\documentclass[lettersize,journal]{IEEEtran}
\IEEEoverridecommandlockouts
\usepackage{cite}
\usepackage{amsmath,amssymb,amsfonts}
\usepackage{algorithmic}
\usepackage{graphicx}
\usepackage{textcomp}
\usepackage{xcolor}
\usepackage[a4paper, total={184mm,239mm}]{geometry}
\usepackage[ruled,vlined]{algorithm2e}
\usepackage{amsmath}
\usepackage{amssymb} 
\usepackage{booktabs}   
\usepackage{multirow}   
\usepackage{array}      
\usepackage{makecell} 
\usepackage{pifont}
\usepackage{enumitem}

\usepackage[table]{xcolor}
\usepackage{tabularx}
\usepackage{booktabs}
\usepackage{array}
\usepackage{caption}
\usepackage[switch]{lineno}
\usepackage{cancel}
\usepackage{xurl}
\usepackage{hyperref} 

\def\BibTeX{{\rm B\kern-.05em{\sc i\kern-.025em b}\kern-.08em
    T\kern-.1667em\lower.7ex\hbox{E}\kern-.125emX}}
    
\newcommand{\bh}[1]{\noindent\textbf{#1}}

\newcommand{\revcam}[1]{\textcolor{black}{#1}}
\newcommand{\rev}[1]{\textcolor{black}{#1}}

\newenvironment{condensed}
{
  \fontdimen2\font=0.9\fontdimen2\font 
  \fontdimen3\font=0.8\fontdimen3\font 
  \fontdimen4\font=0.8\fontdimen4\font 
}

\newboolean{RevResponse}
\setboolean{RevResponse}{false} 
\newboolean{setAppendix}
\setboolean{setAppendix}{false}
\newboolean{CamResponse}
\setboolean{CamResponse}{false} 

\title{HYDRA: A Heterogeneous Chiplet DSE Framework for Serving Dynamic Hybrid LLM Workloads}

\author{%
Jiahao~Lin\textsuperscript{1},
Alish~Kanani\textsuperscript{1},
Sangwan~Lee\textsuperscript{2},
Jaehyun~Park\textsuperscript{2},
and Umit~Y.~Ogras\textsuperscript{1}
\\[-0.15em]
{\normalfont\normalsize
\textsuperscript{1}University of Wisconsin--Madison
\quad
\textsuperscript{2}University of Ulsan}
\\[-0.15em]
{\normalfont\normalsize
\{jlin445,ahkanani,uogras\}@wisc.edu
\quad
\{lso500,jaehyun\}@ulsan.ac.kr}%
\thanks{This work was supported by the R\&D Promotion Foundation for Special Zones funded by the Ministry of Science and ICT of Korea (Grant No. 2025-0057) and the ANCHOR program through the Ulsan ANCHOR Center, funded by the Ministry of Education and the Ulsan Metropolitan City, Republic of Korea (2026-ANCHOR-07-001).
(Corresponding authors: Jaehyun Park and Umit Y. Ogras)}
}

\begin{document}
\IEEEaftertitletext{\vspace{-2.4\baselineskip}}


\maketitle

\begin{abstract}
Hybrid Transformer–Mamba large language models (LLMs) enhance long-context efficiency, but their heterogeneous computation and communication patterns complicate efficient hardware acceleration. Chiplet-based architectures offer a scalable solution by integrating specialized compute and memory units. 
However, the design space spanning static architectural configurations and dynamic runtime policies is prohibitively large to explore exhaustively.
To address this challenge, we present HYDRA, a comprehensive design space exploration framework for hybrid LLM serving on heterogeneous chiplet systems. 
HYDRA jointly explores chiplet composition, placement, inter-chiplet bandwidth provisioning, dynamic batching, and runtime scheduling. 
It integrates communication-aware placement, dynamic batching, elastic task scheduling, and a fast Markov–based performance estimator that captures multi-tenant runtime dynamics for efficient and accurate exploration.
\revcam{Across all workloads, HYDRA delivers 1.55$\times$ the throughput and 43.7\% lower time-to-first-token on average, with throughput gains reaching up to 2.3$\times$,} compared to state-of-the-art baselines.
These results highlight that co-designing architecture and runtime policies is critical for efficient large-scale LLM serving on heterogeneous chiplet systems.

\end{abstract}


\section{Introduction}\label{sec:intro}




The rapid growth of large language models (LLMs) is exposing fundamental limitations in existing hardware systems for efficient high-throughput, low-latency serving. 
Transformer-based models remain the dominant foundation and are widely deployed on graphics processing units (GPUs), tensor processing units (TPUs), and accelerators tailored for autoregressive workloads. 
State-space models (SSMs), including S4~\cite{gu2022efficiently} and Mamba~\cite{gu2024mamba}, have recently emerged as compelling alternatives by offering linear computational complexity and competitive performance for long-context inference. 
\rev{Unlike self-attention, whose KV-cache storage and memory traffic grow with context length, SSM layers maintain a compact recurrent state and provide linear-time sequence processing. As a result, SSMs substantially reduce memory pressure during long-context inference, making them increasingly attractive for high-throughput LLM serving.}
Their sequential dataflow and reduced memory footprint have motivated specialized accelerator micro-architectures~\cite{li2024marca,wei2025lightmamba,kim2025emamba}. 
Building on these advances, recent LLMs increasingly adopt hybrid Transformer–Mamba architectures (e.g., Jamba~\cite{lenz2025jamba}, Nemotron-H~\cite{blakeman2025nemotron}, and Zamba~\cite{glorioso2024zamba}). These new models introduce heterogeneous execution patterns within a single model by combining the scalable long-context processing of SSMs with the strong dependency modeling of transformers.

Accelerating hybrid LLMs is challenging due to the heterogeneity across their model components and runtime serving phases. On the model side, SSM blocks exhibit a growing proportion of element-wise operations as sequence length grows~\cite{li2024marca}. 
In contrast, Transformer computations are dominated by matrix multiplications~\cite{amirshahi2024accelerator}. 
Mamba further introduces nonlinear operators, such as gating and activation functions, which increase
control complexity and hardware area.
The prefill and decode phases further amplify this heterogeneity~\cite{zhong2024distserve,kanani2026duet}. The prefill stage, especially for long contexts, is compute-intensive for both Transformer and Mamba blocks due to the high degree of parallel token processing. 
During decoding, however, the two model components exhibit fundamentally different behaviors. Mamba blocks sustain nearly constant compute and memory demand regardless of context length, whereas Transformer blocks incur linearly increasing memory and bandwidth overhead due to KV-cache access.
These model- and phase-dependent behaviors necessitate the \textit{co-design of heterogeneous chiplet architectures and runtime scheduling policies to effectively exploit workload characteristics.}

\begin{figure}[t]
\centering
    \centering
    \includegraphics[width=1\linewidth]{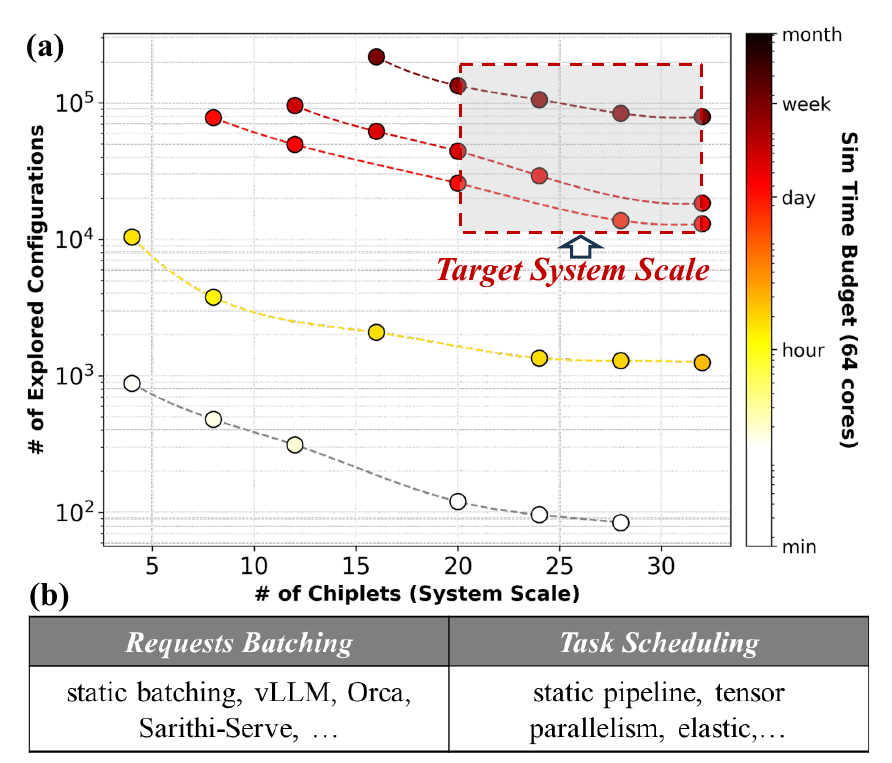}
    \caption{(a) Design space of the chiplet-based system's macroarchitecture and corresponding simulation cost (b) Runtime and design-time scheduling strategies for LLM services.}
    \label{fig:hybrid_profile}
    \vspace{-6mm}
\end{figure}

The integration of diverse processing units onto a single monolithic chip for large-scale LLM workloads has become increasingly cost-prohibitive~\cite{sharma2025heterogeneous}. Consequently, heterogeneous chiplet-based architectures have emerged as a promising solution for accelerating multi-tenant hybrid LLM services by enabling modular integration of compute, memory, and networking components~\cite{xu2025wsc}. 
However, designing chiplet-based systems for hybrid LLM serving requires the joint co-design of architecture and runtime scheduling, significantly expanding the design space.
Specifically, \textit{macro-architecture} design space exploration (DSE) for large-scale heterogeneous platforms introduces two key challenges:

\bh{(I) Broad Design Space:} Heterogeneous architectures for hybrid LLM workloads require heterogeneous chiplets optimized for Transformer
and Mamba computations, each further tailored to prefill and decode phases. 
In addition, the area allocated to compute chiplets must be balanced against memory capacity and inter-chiplet communication resources. 
The number of candidate configurations grows rapidly with both system scale and architectural heterogeneity. 
Fig.~\ref{fig:hybrid_profile}(a) shows the maximum number of configurations that can be explored under a given simulation time budget. Even with fixed chiplet placement and scheduling policies, exhaustive simulation of large-scale heterogeneous systems can require days to weeks of wall-clock time on a 64-core CPU~\cite{amd_threadripper_aec}. 
Therefore, naïve DSE becomes impractical, motivating the need to reduce exploration time from weeks to minutes. 

\bh{(II) Runtime Unpredictability:}
Auto-regressive decoding depends on input-dependent behavior, while multi-tenant serving introduces asynchronous request arrivals with diverse prefill and decode characteristics~\cite{hu2024blockllm}. These factors cause highly dynamic resource demands across computation, memory, and inter-chiplet communication, limiting the effectiveness of offline task mapping and static performance models commonly used in DSE frameworks. As illustrated in Fig.~\ref{fig:hybrid_profile}(b), modern LLM serving relies on dynamic batching (e.g., vLLM~\cite{kwon2023efficient}, Orca~\cite{yu2022orca}, Sarathi-Serve~\cite{agrawal2023sarathi}) and diverse task scheduling strategies, necessitating runtime-aware modeling for accurate evaluation. 
\textit{Existing DSE frameworks for chiplet-based systems largely assume static workloads and fixed scheduling policies}, and therefore fail to capture the dynamic, multi-tenant behavior of hybrid LLM serving~\cite{cai2024gemini,adiletta2025democratizing,xu2025wsc}.

To address these challenges, we propose \textit{HYDRA}, a DSE framework for heterogeneous chiplet-based systems targeting hybrid LLM serving. \textit{HYDRA} integrates communication-aware chiplet placement, dynamic request batching, and elastic task scheduling with a high-level simulator and a Markov-based performance estimator that guide the efficient and accurate DSE. 
\rev{The Markov-based estimator is not intended to replace detailed simulation. Instead, it serves as a fast pruning and ranking mechanism that rapidly identifies promising regions of the design space. We complement it with detailed simulations to validate selected configurations and accurately capture transient contention, NoI communication effects, memory-management behavior, dynamic batching decisions, and runtime pipeline imbalance that are intentionally abstracted by the estimator.}
Together, these components capture runtime variability and enable efficient co-exploration of architecture and runtime policies.
\textit{HYDRA} identifies Pareto-optimal configurations in the throughput and time-to-first-token (TTFT) space. 
Across all evaluated workloads, \revcam{\textit{HYDRA} delivers 1.55$\times$ the throughput and 43.7\% lower TTFT on average.} To accelerate exploration, we develop a Markov-based performance estimator that characterizes runtime transitions among chiplet states and predicts system throughput. This estimator achieves an average cosine similarity of 0.9 with full simulation results and reduces DSE time from days to minutes.

\textit{Main contributions of this work are:}
\begin{itemize}
    \item An DSE framework for hybrid LLM workloads on 2.5D heterogeneous chiplet systems, available at \href{https://github.com/ONQLin/hydra-cases-2026-artifact/tree/v1.0.0-cases2026}{Github}.
    \item A communication-aware chiplet placement strategy, combined with dynamic batching and elastic task scheduling, to improve compute and memory utilization,
    \item A Markov-based performance estimator for hybrid LLM workloads that enables fast and accurate DSE,
    \item Comprehensive evaluations across hybrid, Mamba, and Transformer models on diverse datasets.
\end{itemize}

\section{Related Work}\label{sec:related_work}

Hybrid Transformer--Mamba LLMs leverage Transformer-style attention for in-context learning with Mamba layers that enable recurrent linear-complexity sequence modeling. 
This fusion improves scalability to long context lengths while reducing memory overhead compared to pure Transformer baselines. 
\rev{Models with a larger Attention fraction tend to favor higher memory capacity and bandwidth due to KV-cache traffic, whereas Mamba-dominated models place greater emphasis on recurrent-state processing and sustained decoding throughput.}
For example, Jamba integrates the hybrid architecture with Mixture-of-Experts (MoE) layers, achieving competitive accuracy with up to 3$\times$ higher throughput~\cite{lenz2025jamba}. Nemotron-H and Minitron-SSM employ a similar hybrid strategy, introducing a unified SSM pruning pipeline that compresses the model by 50\% with minimal quality impact~\cite{blakeman2025nemotron,taghibakhshi2026efficient}.

Significant research has focused on optimizing the prefill and decode phases of Transformer and Mamba accelerators.
Dedicated accelerators have been proposed for either transformer (e.g., TSTC~\cite{liu2023tstc}, and AccelTran~\cite{tuli2023acceltran}) or Mamba (e.g., LightMamba~\cite{wei2025lightmamba}, SSM-RDU~\cite{ko2025ssm}, eMamba~\cite{kim2025emamba}, and SpecMamba~\cite{zhong2025specmamba}). 
Recent approaches propose unified architectures that efficiently support both SSM computation and matrix multiplication. For example, MARCA introduces a reconfigurable tensorcore-like architecture that accelerates element-wise and specific nonlinear operations~\cite{li2024marca}. 
Geens et al. further explore MARCA’s design space and reveal how different allocations of compute and on-chip memory resources shift bottlenecks across serving phases~\cite{geens2025fine}. 
While these works focus on accelerator micro-architecture,
\textit{HYDRA focuses on system-level macro-architecture DSE for hybrid LLM serving}. 
Unlike prior work, our design space explicitly captures heterogeneous accelerator configurations across both prefill and decode phases.
These observations highlight the need for system-level frameworks that jointly optimize how heterogeneous compute and memory resources are composed, placed, and scheduled in chiplet-based systems for end-to-end LLM serving.

Determining the optimized composition of heterogeneous chiplets for hybrid LLM workloads remains an open problem. 
Prior DSE frameworks have explored different parts of the design space of chiplet-based systems. Gemini adopts a scalable Simba-like architecture and performs layer-wise pipeline mapping for DNN workloads, targeting homogeneous compute chiplets~\cite{cai2024gemini}. 
Focusing on edge deployments, Cascade clusters multi-tenant workloads based on their compute and memory characteristics, and assigns them across chiplet packages~\cite{adiletta2025democratizing}. 
WSC-LLM targets LLM workloads deployed on wafer-scale chips and explores micro-architectural trade-offs across DRAM capacity, communication bandwidth, and compute resources distributed across chiplets~\cite{xu2025wsc}. It also proposes a disaggregated scheduling strategy that separates prefill and decode execution, an approach also used in modern GPU serving systems. 
\textit{Despite these advances, existing DSE frameworks largely assume static workload traces and do not capture the dynamic, multi-tenant behavior of LLM serving} (e.g., varying prefill–decode lengths and asynchronous request arrivals). As summarized in Table~\ref{tab:chiplet_dse_compact}, \textit{HYDRA} addresses both runtime dynamics and macro-architecture exploration, which are essential for efficient hybrid LLM serving on heterogeneous chiplet systems. 

\begin{table}[t]
\centering
\caption{Comparison of prior chiplet-based DSE frameworks across workload support, heterogeneity, scalability, target design space, and runtime-dynamics coverage.}
\vspace{0mm}
\renewcommand{\arraystretch}{1.05}
\setlength{\tabcolsep}{1pt}
\resizebox{\linewidth}{!}{%
\begin{tabular}{l|c|c|c|c|c}
\toprule
\textbf{Framework} &
\textbf{Workload} &
\textbf{Hetero.} &
\textbf{Scale} &
\textbf{Target DS} &
\textbf{Runtime Dyn.}
\\
\midrule

Gemini &
CNN/MLP &
$\times$ &
8 chiplets &
Microarch. &
$\times$ 
\\

Cascade &
Edge App.&
$\checkmark$ &
12 chiplets &
Macroarch. &
$\times$
\\

WSC-LLM &
Transformer LLM&
$\times$ &
64 chiplets &
Microarch. &
$\times$
\\

\midrule
\textbf{\textit{HYDRA}} &
\textbf{Hybrid LLM} &
$\pmb\checkmark$ &
\textbf{24 chiplets} &
\textbf{Macroarch.} &
$\pmb\checkmark$ 
\\
\bottomrule
\end{tabular}
}
\vspace{-4mm}
\label{tab:chiplet_dse_compact}
\end{table}

Recent LLM evaluation and scheduling frameworks also aim to capture LLM-specific dynamics in system behavior. For example, Vidur is a simulation framework for LLM inference on GPUs that integrates a workload generator with various runtime schedulers to model realistic multi-request serving scenarios~\cite{agrawal2024vidur}. 
Our work complements these efforts by coupling hybrid LLM workload modeling with chiplet-aware DSE and runtime scheduling co-design, enabling a holistic analysis of performance bottlenecks and system-level trade-offs in heterogeneous chiplet architectures.

\vspace{-0.5mm}
\section{Hydra DSE Framework}
\vspace{-1mm}

\subsection{Overview}
\begin{figure}[t!]
\centering
    \centering
    \includegraphics[width=0.95\linewidth]{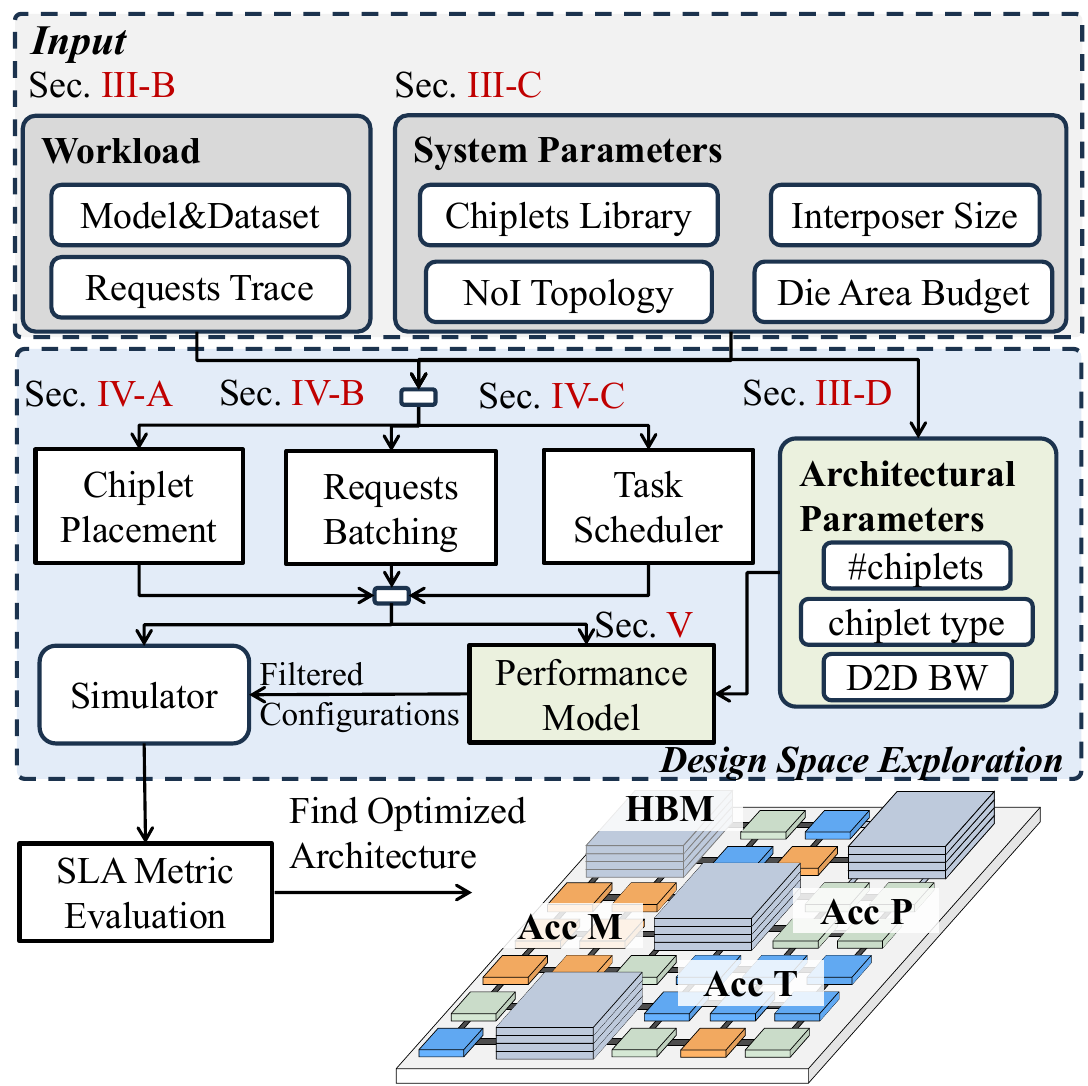}
    \caption{The overview of the proposed design space exploration framework targeting hybrid LLM services.}
    \label{fig:overview}
\vspace{-6mm}
\end{figure}

The \textit{HYDRA} framework takes two classes of inputs, as illustrated in Fig.~\ref{fig:overview}: 
(1) the workload specification, including the target models, datasets, and request traces, 
(2) system constraints, including the chiplet library, network-on-interposer (NoI) topology, interposer size, and die-area budget. 
Given these inputs, \textit{HYDRA} performs joint exploration across four architectural and runtime dimensions:
\begin{enumerate}
    \item chiplet placement (Section~\ref{sec:chip_placmt}), 
    \item request batching policy (Section~\ref{sec:dynamic_batch}), 
    \item task scheduling strategy (Section~\ref{sec:task_sched}), 
    \item the macro-architecture, including chiplet types and quantities (Section~\ref{sec:dse_intro}).
\end{enumerate}

Each candidate configuration is evaluated using a combined simulation and performance estimation flow that captures prefill/decode behaviors, multi-tenant concurrency, and inter-chiplet communication. 
Feasible configurations are compared in terms of throughput and TTFT. This modular structure enables \textit{HYDRA} to capture the interactions between architecture, runtime scheduling, and workload dynamics during exploration.

Overall, \textit{HYDRA} models hybrid LLM serving as a tightly coupled system of (i) workload dynamics, (ii) heterogeneous chiplet resources, and (iii) runtime batching/scheduling policies. The DSE process jointly explores architectural configurations and runtime decisions under system constraints to optimize system-level throughput and TTFT.

\vspace{-2mm}
\subsection{Workload Description}\label{sec:workloads}

We model hybrid LLMs serving workloads with diverse computational and memory behaviors
across application domains. 
The proposed workload model captures both structural heterogeneity (across model components) and temporal dynamics (across requests and serving phases), as described next.

\noindent\textbf{Models:}
\textit{HYDRA} takes a layer- or block-level description, where each block is annotated with its compute, memory, and communication requirements.
This representation supports heterogeneous LLMs composed of different operator types and execution patterns, including Attention-based blocks ($A$), Mamba-based blocks ($M$), or their combinations. 
Example models used for evaluation are described in Section~\ref{sec:setup}.

\noindent\textbf{Datasets:} 
We employ commonly used benchmarks that cover diverse inference patterns, including summarization, translation,
interactive chat, and long-form generation, 
each with distinct prefill/decode characteristics. \textit{HYDRA} also supports a configurable request arrival process to control the relative intensity of prefill and decode.

\noindent\textbf{Service dynamics:}
Each request undergoes a \textit{prefill phase}, where the full input sequence is processed once, followed by a \textit{decode phase}, where tokens are generated autoregressively. 
In multi-tenant scenarios, decode-phase duration and resource requirements are input-dependent, while chiplet utilization varies dynamically as requests arrive and complete. 
These temporal variations drive the compute, memory, and communication demands, making runtime scheduling and batching essential for efficient system utilization.

\noindent\textbf{Task Definition:}
\textit{HYDRA} translates workloads into task-level execution streams used for both simulation and performance estimation.
Each task $k$ represents all compute and memory operations of a model operation during inference, such as a fully connected layer, an Attention, or SSM operation. 
$\mathcal{K}$ denotes the set of all schedulable tasks generated during inference for the target model.

\vspace{-3mm}
\subsection{System Specification}\label{sec:systemspec}
The system configuration is specified by four aspects: the chiplet library, the NoI, the interposer size, and the die area budget. The chiplet library and NoI determine the system’s architectural capabilities and heterogeneity, while the interposer size and die-area budget impose physical and cost constraints on feasible designs.

\noindent\textbf{Chiplet Library:} 
\textit{HYDRA} maintains a library of compute and memory chiplets. Compute chiplets may differ in accelerator organization, phase specializations (e.g., optimized for prefill or decode), and target model component (e.g., optimized for Transformer or SSM blocks). 
This abstraction enables \textit{HYDRA} to model a broad range of candidates, including commercial-inspired designs (e.g., B200-like~\cite{nvidia_blackwell_datasheet_2024}), prior accelerator architectures (e.g., TSTC~\cite{liu2023tstc} and MARCA~\cite{li2024marca}), and customized systolic-array- or vector-based architectures. 
Memory chiplets may vary in technology, capacity, and bandwidth, such as \rev{High Bandwidth Memory (HBM3~\cite{park2022192})} and GDDR7~\cite{jedec2024jesd239a}. 

\noindent\textbf{NoI Specification:} 
\textit{HYDRA} models the NoI, including topology (e.g., mesh and torus~\cite{kannan2015enabling}), die-to-die (D2D) bandwidth, latency, and communication cost. These parameters determine the efficiency of data movement across chiplets, including weights, activations, KV cache, and recurrent states. 

\begin{figure}[t]
\centering
    \centering
    \includegraphics[width=1\linewidth]{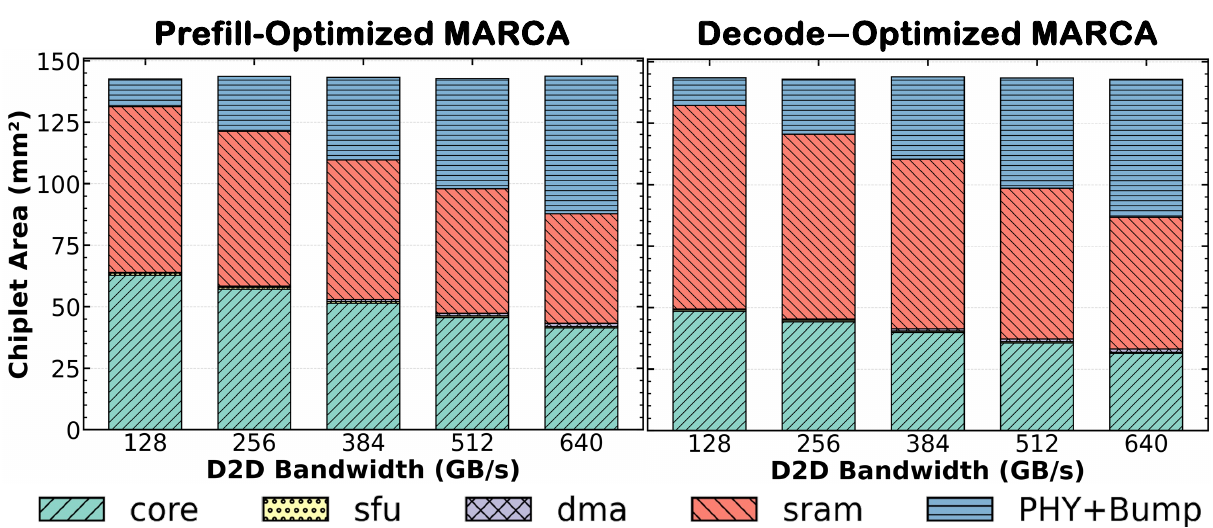}
    \caption{The area breakdown of compute chiplets in different NoI bandwidth. Cores and SFUs are compute units for linear and non-linear operations. SRAM represents local buffers. DMA, PHY, and bumps represent communication components.}
    \label{fig:comp_breakdown}
\vspace{-5mm}
\end{figure}

D2D bandwidth provisioning also affects chiplet microarchitecture under a fixed die-area budget. Increasing bandwidth requires additional communication hardware, such as PHYs, micro-bumps, and routing logic, which reduces the area available for compute units and on-chip SRAM. Consequently, bandwidth, computation, and local buffering must be co-optimized during DSE. 
For illustration, Fig.~\ref{fig:comp_breakdown} shows how D2D bandwidth provisioning affects the area breakdown of MARCA~\cite{li2024marca} chiplets optimized for prefill and decode.
\textit{These trade-offs directly influence the optimal chiplet composition and placement decisions explored during DSE.}

\noindent\textbf{Interposer Size:} 
\textit{HYDRA} uses the interposer size to constrain the physical integration space of the chiplet system. This parameter bounds the number of chiplets that can be integrated and affects their feasible placement, shaping the architectural design space explored by the framework. The concrete interposer assumptions used in evaluation are given in Section~\ref{sec:setup}.

\noindent\textbf{Die Area Budget:} 
Each chiplet in \textit{HYDRA} is subject to a die-area budget that limits the amount of compute, on-chip memory, and communication hardware that can be integrated. Together with the NoI specification, this budget determines the trade-off among compute resources, SRAM capacity, and communication interfaces within each chiplet. 

\definecolor{cfghead}{RGB}{214,228,248}   
\definecolor{deshead}{RGB}{217,239,224}   
\definecolor{modhead}{RGB}{252,227,206}   
\definecolor{cfgrow}{RGB}{214,228,248}    
\definecolor{desrow}{RGB}{217,239,224}    
\definecolor{modrow}{RGB}{252,227,206}    

\newcolumntype{L}[1]{>{\raggedright\arraybackslash}p{#1}}
\newcolumntype{Y}{>{\raggedright\arraybackslash}X}

\begin{table}[b!]
\centering
\caption{
Summary of
notations used in this work
}
\label{tab:notation}
\setlength{\tabcolsep}{3pt}
\renewcommand{\arraystretch}{1.03}
\footnotesize

\begin{minipage}[t]{0.487\textwidth}
\centering
\begin{tabularx}{\linewidth}{@{}L{0.2\linewidth}X@{}}
\toprule
\textbf{Notation} & \textbf{Definition} \\
\midrule

$k \in \mathcal{K}$ &
Schedulable task instance ($k$) and set of tasks ($\mathcal{K}$). \\

$A_p,A_d$ &
Prefill-/decode-optimized ($A_p$/$A_d$) Attention accelerators\\

$M_p,M_d$ &
Prefill-/decode-optimized ($M_p$/$M_d$) Mamba accelerators\\

$C_A$ &
The set of Attention accelerator chiplets \\

$C_M$ &
The set of Mamba accelerator chiplets \\

$Loc_M, Loc_A$ &
The set of locations of HBM chiplets allocated to the Mamba blocks ($Loc_M$) or Attention blocks ($Loc_A$) \\

$r_M,r_A$ &
The maximum request concurrency supported by the Mamba HBM group ($r_M$) or Attention HBM group ($r_A$) \\

$v(k,c)$ &
Communication volume of task $k$ on chiplet $c$. \\


$p(c)$ &
Chiplet placement location of $c$\\

$t_{comp}(k,c)$ &
Compute latency of task $k$ on chiplet $c\in C_A \cup C_M$ \\

$t_{comm}(k,c)$ &
Communication latency of task $k$ on chiplet $c$ \\

$\Omega(k)$ &
Task assignment of $k$ to a chiplet\\

\bottomrule
\end{tabularx}
\end{minipage}












\end{table}

\vspace{-2mm}
\subsection{Design Space}
\label{sec:dse_intro}

This section defines the macro-architecture design space explored by \textit{HYDRA}.
It captures how compute, memory, and communication resources are provisioned at the package level.
An architectural configuration is defined by three key aspects:
\begin{itemize}[leftmargin=*]
    \item \textit{Chiplet types}: The set of available chiplet types, including prefill-optimized Attention ($A_p$), decode-optimized Attention ($A_d$), prefill-optimized Mamba ($M_p$), and decode-optimized Mamba ($M_d$) accelerators, as summarized in Table~\ref{tab:notation}. 
      
    \item \textit{Chiplet composition}: The number of instantiated chiplets for each type, which determines the overall mix of compute and memory resources in the system. For example, a composition defines the instantiated accelerator sets $C_A$ and $C_M$ for Attention and Mamba, respectively.

    \item \textit{NoI bandwidth}: The D2D bandwidth provisioning under a given topology (e.g., 128 or 256~GB/s links).

\end{itemize}

These architectural choices are constrained by the interposer area budget, i.e., the total area of all instantiated chiplets must not exceed the available interposer area.
Under a given topology and area constraint, each architectural configuration is defined by the selected chiplet types, their composition, and D2D bandwidth. 
For each such configuration, \textit{HYDRA} further explores runtime and physical design decisions, including chiplet placement, dynamic request batching, and elastic task scheduling (Section~\ref{Sec:Opt}). 

Overall, \textit{HYDRA} jointly explores architectural configurations and runtime policies to capture their interactions and identify designs that optimize throughput and TTFT.
Each candidate configuration can be evaluated via full simulation, which models dynamic request arrivals, batching behavior, and elastic scheduling. However, as shown in Fig.~\ref{fig:hybrid_profile}, simulation becomes prohibitively time-consuming for large-scale systems and dynamic LLM workloads. Therefore, we use a fast and accurate performance model, introduced in Section~\ref{sec:markov}.

\section{LLM Inference Optimization Strategies}\label{Sec:Opt}
\subsection{Chiplet Placement}\label{sec:chip_placmt}

\textit{HYDRA} models chiplet placement on a 2D grid-based interposer layout, where each grid cell represents the minimum allocatable unit. A chiplet may occupy one or multiple adjacent cells depending on its physical footprint, enabling us to capture diverse chiplet sizes and layout constraints. For example, each cell can be 5 mm x 5 mm, and the chiplet of size 100 mm$^2$ can occupy four of these cells. This abstraction supports a flexible representation of candidate placements.

\begin{figure}[h]
\vspace{-3mm}
\centering
    \centering
    \includegraphics[width=1\linewidth]{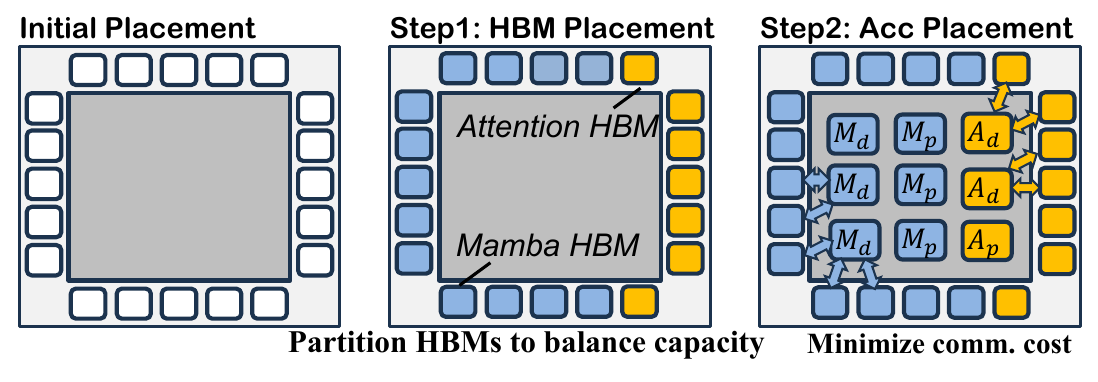}
    \vspace{-5mm}
    \caption{The overview of the chiplet placement in \textit{HYDRA}}
    \label{fig:Placement}
    \vspace{-2mm}
\end{figure}

Given the chiplet composition, the placement objectives are: 
(i) maximize the number of concurrent requests processed under memory constraints, and (ii) minimize inter-chiplet communication cost. 
These objectives are tightly coupled and depend on workload characteristics, making joint optimization intractable for large design spaces.
To make this problem tractable, we decompose placement into two stages that separately address memory capacity constraints and communication locality, as illustrated in Fig.~\ref{fig:Placement}.

\noindent\textbf{Step 1: HBM Placement.}
HBM chiplets are placed along the outer ring of the interposer 
to reflect their role as high-bandwidth memory endpoints with efficient access to package-level interfaces and power delivery.
This organization is common in prior chiplet-based systems~\cite{amd_mi350_series_2025,nvidia_blackwell_datasheet_2024}. 
Given these locations, we partition the HBM chiplets into Mamba and Attention groups to provision memory for their respective workloads.
Since these workloads exhibit distinct memory-access patterns and runtime bottlenecks, separating memory resources reduces interference and improves utilization.

For a candidate partition $(N_M, N_A)$, where $N_M$ and $N_A$ denote the number of HBM chiplets assigned to Mamba and Attention, respectively, we estimate the maximum concurrent requests supported by each group.
We first estimate the maximum concurrency supported by the Mamba HBM group, denoted by $r_M$, as:
\begin{equation}
r_M=\max\left(0,\left\lfloor
\frac{N_M \cdot C_{HBM}- L_M \cdot W_M}{S_M}
\right\rfloor\right) 
\end{equation}
where $C_{HBM}$ is the capacity of a single HBM, $L_M$ is the number of Mamba blocks in the model, $W_M$ is the weight memory required by one Mamba block, and $S_M$ is the per-request memory required to store the Mamba recurrent states. 
Here, $N_M \cdot C_{HBM}$ denotes the total HBM capacity assigned to Mamba, while $L_M \cdot W_M$ denotes the memory reserved for all Mamba weights. 
Therefore, the numerator is the remaining HBM capacity available for runtime state allocation, and $r_M$ gives the maximum concurrency supported by the Mamba HBM group.

The maximum concurrency for the Attention HBM group, denoted by $r_A$, is estimated in the same manner by replacing the Mamba weight and recurrent-state terms with the corresponding weights and KV-cache terms. 
Then, we select the partition ($N_M, N_A$) that maximizes $\min(r_M, r_A)$ to avoid memory bottlenecks in either component. 
Once the best partition is found, the corresponding outer-ring HBM locations are labeled as the Mamba HBM group and the Attention HBM group, and are denoted $Loc_M$ and $Loc_A$, respectively. These two sets of locations are then passed to Step 2.


\noindent\textbf{Step 2: Communication-Aware Placement.} 
Given the HBM locations from Step 1, compute chiplets are placed to minimize the inter-chiplet communication overhead.
The objective is to place each compute chiplet close to the memory resources it accesses most frequently.

Let $C_M$ and $C_A$ denote the sets of Mamba and Attention compute chiplets 
(defined in Table~\ref{tab:notation}).
Thus, $C_M \cup C_A$ constitutes the full set of compute chiplets to be placed.
Consider the placement of Mamba compute chiplets $C_M$ for illustration. 
We first select candidate locations for all Mamba compute chiplets based on the smallest Manhattan distance to $Loc_M$.
\textit{For a given task $k \in \mathcal{K}$}, the objective is to find the placement that minimizes the total weighted communication cost:
\begin{equation}
\arg\min_{p}~\sum_{c\in C_M}\sum_{h\in Loc_M} d(p(c), h)\cdot v(k,c)
\end{equation}
where $p(c)$ denotes the location of chiplet $c$, $d(\cdot,\cdot)$ is the Manhattan distance between two locations, and $v(k,c)$ is the communication volume required when task $k$ is executed on chiplet $c$. For example, $k$ denotes the state-space model~\cite{gu2022efficiently} computation when placing the Mamba chiplets $C_M$.


The placement of the Attention chiplets in $C_A$ is determined in the same manner using $Loc_A$. 
Because $C_M$ and $C_A$ are placed at the same grid-based interposer layout, their preferred locations may overlap. 
To resolve location conflicts, we place $C_M$ and $C_A$ sequentially in descending order of their sizes, allowing larger groups to claim preferred locations first. 
After one group is placed, its assigned locations are removed from the available set before constructing the candidate location set for the other group. 
Overall, this strategy co-locates compute chiplets with their dominant communication partners, reducing NoI traffic and latency.
Section~\ref{sec:placmt} evaluates the effectiveness of this placement strategy.

In summary, the proposed two-stage decomposition separates memory capacity balancing from communication optimization, enabling efficient exploration while capturing the dominant performance bottlenecks in hybrid LLM serving.

\subsection{Dynamic Request Batching}\label{sec:dynamic_batch}



\begin{figure}[t]
\centering
    \centering
    \includegraphics[width=1\linewidth]{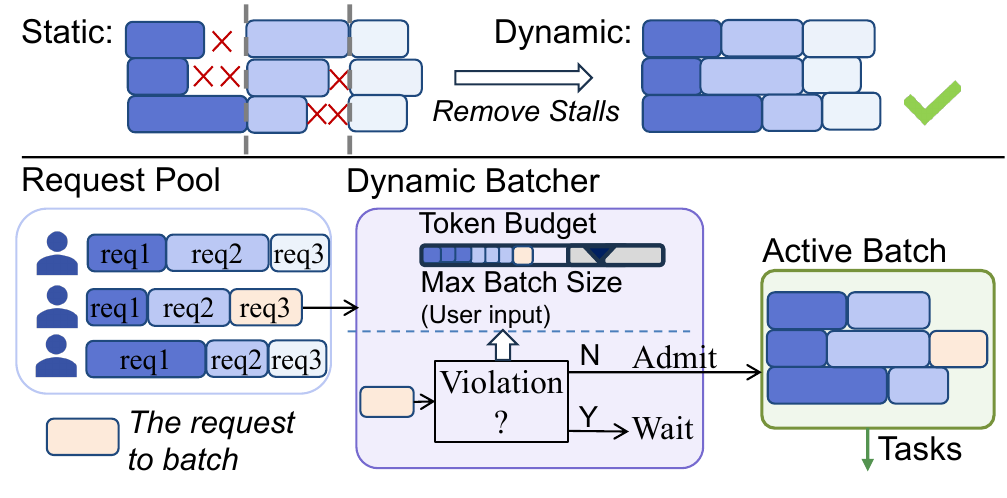}
    \vspace{-5mm}
    \caption{The overview of the \textit{dynamic batching} in \textit{HYDRA}}
    \label{fig:Runtime_b}
\vspace{-5mm}
\end{figure}

Request batching is a key optimization in LLM serving because it improves parameter reuse and exposes token-level parallelism. 
However, in multi-tenant scenarios, requests arrive asynchronously and exhibit highly variable execution costs: input lengths differ, decode lengths are input-dependent, and resource demand cannot be known a priori. 
Under these conditions, static batching is inefficient, as illustrated in the top part of Fig.~\ref{fig:Runtime_b}.
Static execution is often gated by the longest request, while shorter requests complete early but cannot be replaced. This leads to stalls and underutilization of compute, memory, and communication resources.

To address this limitation, \textit{HYDRA} adopts dynamic request batching, maintaining a continuously evolving active batch rather than fixed batch boundaries. 
As shown in Fig.~\ref{fig:Runtime_b}, requests are admitted as soon as resources become available, without waiting for batch-level synchronization. 
Incoming requests first enter a request pool with varying input sizes and memory requirements. 
A request is admitted into the active batch if two conditions are met:
\begin{enumerate}
    \item The active batch size is smaller than a user-defined maximum since larger batches can significantly degrade per-request prefill latency, as evaluated in Section~\ref{sec:dse_res},
    \item Adding the new request does not violate the available token budget.
\end{enumerate}
\rev{We compute a token budget as the product of the current batch size and median per-request token cost.
New requests are admitted only when the estimated token demand remains below a system-level token budget derived from available compute, memory, and communication resources.}

To enable continuous admission without the overhead of frequent reallocation, memory management is decoupled from request admission via virtualized memory blocks, similar to vLLM~\cite{kwon2023efficient}. 
Each request is assigned virtual memory for its execution state, which grows dynamically with KV-cache or recurrent state during decoding, while physical memory is allocated on demand. 
This design avoids large contiguous allocations and improves robustness to variable request footprints.

\textit{HYDRA} employs request-level preemption to prevent out-of-memory conditions. 
\rev{If admitting a new request or expanding an existing request's KV/state storage would exceed the available HBM capacity, \textit{HYDRA} preempts the request with the fewest generated tokens, reclaims its memory allocation, and returns it to the request pool. This policy prioritizes requests that are closer to completion, reducing wasted work while maintaining forward progress under memory pressure.
Whenever requests are admitted, completed, or preempted, the active batch is updated accordingly. The resulting batch snapshot is then passed to the elastic task scheduler for execution.}


Overall, \textit{HYDRA}'s dynamic batching policy generalizes continuous batching to heterogeneous chiplet systems and phase-dependent hybrid LLM workloads, explicitly incorporating trade-offs among compute, memory, communication, and latency into admission control.

\subsection{Elastic Task Scheduler}\label{sec:task_sched}

After request admission, \textit{HYDRA} dispatches schedulable tasks $k \in \mathcal{K}$ to heterogeneous compute chiplets. 
\revcam{
We select the source of $t_{\mathrm{comp}}(k,c)$ according to the chiplet being modeled. In the evaluation (Section~\ref{sec:eval}), Mamba prefill and decode latencies are derived from the microarchitectural parameters and kernel-level characterizations reported for MARCA~\cite{li2024marca}, while Attention prefill and decode latencies are derived from those reported for TSTC~\cite{liu2023tstc}.}
For commercial chiplets such as B200, \textit{HYDRA} derives latency from vendor-published specifications~\cite{nvidia_blackwell_datasheet_2024}, whereas custom designs are characterized through cycle-level RTL simulation~\cite{kanani2026duet}.
These values
form a task--chiplet latency table that captures the affinity between each
task type and compatible chiplet. The profiled latency also includes
chiplet-local memory access under the configured external bandwidth.

\rev{
We estimate $t_{\mathrm{comm}}(k,c)$ using the following equation:
\begin{equation}
t_{\mathrm{comm}}(k,c)=
d_{\mathrm{Man}}(c,m_k)\frac{V_k}{B_{\mathrm{NoI}}}
\end{equation}
where $V_k$ is the data volume transferred by the task,
$d_{\mathrm{Man}}(c,m_k)$ is the Manhattan distance to its memory chiplet, and
$B_{\mathrm{NoI}}$ is the provisioned per-link NoI bandwidth. Thus, the Manhattan distance weights
the transfer cost by the number of links along the path. 
\revcam{A larger transfer occupies NoI links for longer, while a longer path consumes link resources across more hops, and higher bandwidth reduces the transfer time. This first-order model therefore provides a lightweight proxy for initial mapping.}
For tasks accessing multiple memory
objects or memory chiplets, these costs are accumulated. HYDRA uses
$t_{\mathrm{comp}}(k,c)+t_{\mathrm{comm}}(k,c)$ to construct the initial
mapping. Runtime contention is subsequently captured by the simulator.
}

\begin{figure}[t]
\centering
    \centering
    \includegraphics[width=0.95\linewidth]{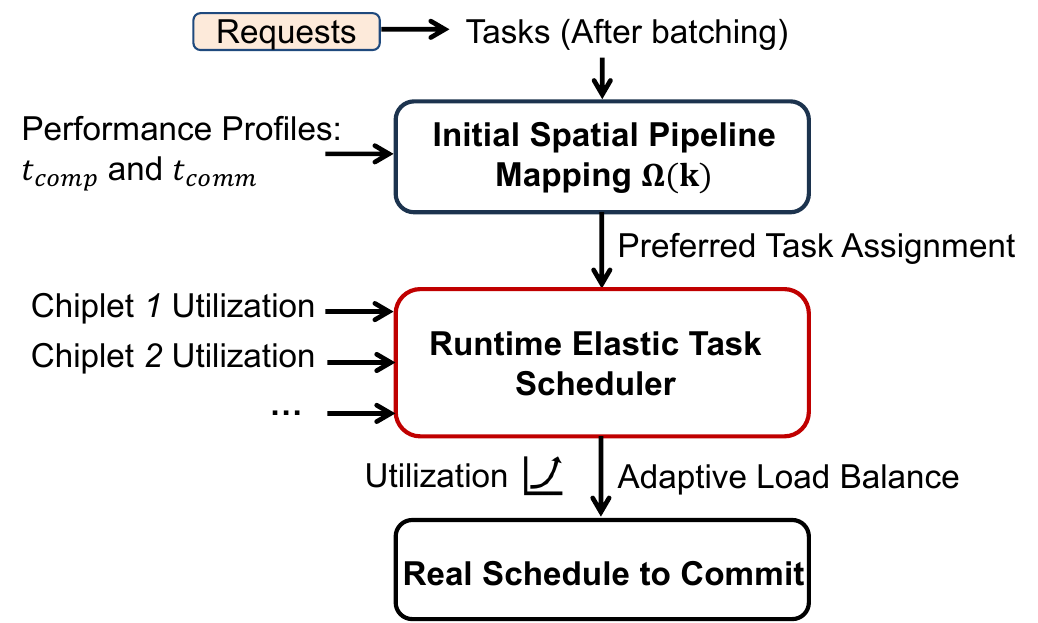}
    \caption{The overview of the task scheduling in \textit{HYDRA}}
    \label{fig:Runtime_t}
    \vspace{-5mm}
\end{figure}

The first objective is to find an initial mapping for each ready task using the performance profiles, as illustrated in Fig.~\ref{fig:Runtime_t}.  
For a given mapping, the cumulative latency assigned to chiplet $c$ can be estimated as:
\begin{equation}
\label{eq:Latency_Chip}
t^{\Omega}_c = \sum_{k : \Omega(k) = c} \left( t_{\text{comp}}(k,c) + t_{\text{comm}}(k,c) \right)
\end{equation}
We aim to minimize the bottleneck stage across chiplets:
\begin{equation}
\label{eq:obj_static}
   \arg\min_{\Omega}~\max_{c \in C_M \cup C_A}~t^{\Omega}_c
\end{equation}
Therefore, \textit{HYDRA} solves this optimization problem to construct an initial spatial pipeline mapping $\Omega(k)$ that stores the preferred chiplet assignment for each task~\cite{cai2024gemini,adiletta2025democratizing}. 
This mapping would be optimal for a static system where the chiplet utilization and ready tasks are known a priori. However, \textit{HYDRA} targets multi-tenant hybrid LLM serving, where the runtime mix of prefilling and decoding requests can vary significantly over time. Under such dynamic conditions, the preferred task assignment may create load imbalance and reduce overall chiplet utilization.
Therefore, \textit{HYDRA} uses a runtime elastic task scheduler to modify the preferred assignments, as illustrated in Fig.~\ref{fig:Runtime_t}. 
The elastic scheduler monitors chiplet utilization through the runtime queue status of compute chiplets. When load imbalance is detected, it adaptively balances the preferred assignments by scheduling more tasks to underutilized chiplets. 

In summary, the \textit{initial} spatial pipeline mapping provides a \textit{preferred task assignment before} execution, while the \textit{runtime elastic task scheduler corrects transient imbalance} caused by dynamic request arrivals and varying prefill/decode pressure. The resulting execution schedule can therefore adapt to runtime conditions without requiring expensive global remapping. This improves chiplet utilization and, in turn, improves throughput and TTFT, as demonstrated in Section~\ref{sec:scheduling}.

\section{Performance Estimation for Fast DSE}
\label{sec:markov}

Evaluating each candidate configuration with full event-driven simulation is prohibitively expensive for large-scale chiplet systems and dynamic LLM workloads. 
\rev{Full simulations must capture asynchronous request arrivals, dynamic batching, elastic scheduling, memory behavior, and NoI contention,
leading to hours of exploration time per design point.} In contrast, static analytical models (e.g., roofline approach~\cite{williams2009roofline}) are fast but fail to capture runtime dynamics, leading to inaccurate performance estimates.
Existing approaches also do not capture the coupling between batching, scheduling, and resource contention, which is critical in multi-tenant LLM serving~\cite{adiletta2025democratizing,cai2024gemini}.
So, there is a strong need for techniques that explicitly incorporate dynamic batching, elastic scheduling, and heterogeneous chiplet capabilities in a unified formulation.

\begin{figure}[t]
    \centering
    \includegraphics[width=1\linewidth]{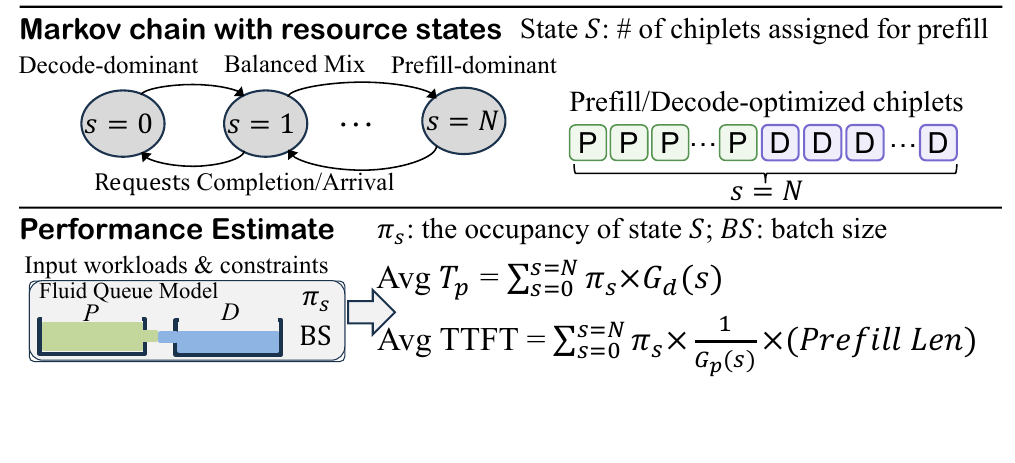}
    \vspace{-10mm}
    \caption{Overview of the Markov-based performance estimation.}
    \label{fig:markov}
    \vspace{-5mm}
\end{figure}

To address this challenge, we develop a lightweight \textit{Markov-based performance estimator} that captures the dominant runtime dynamics while enabling fast DSE, as illustrated in Fig.~\ref{fig:markov}.
\rev{The key abstraction is reducing the high-dimensional scheduling problem into a low-dimensional state space defined by chiplet allocation. It is based on the observation that, hybrid LLM serving performance is primarily governed by how resources are split between prefilling and decoding rather than by the exact ordering of individual tasks. Dynamic batching changes the amount of active work in each phase, while elastic scheduling changes which chiplets serve that work. Therefore, the dominant effect of runtime control can be captured by tracking aggregate chiplet allocation across execution phases.}

\vspace{-3mm}
\subsection{Modeling Elastic Chiplet Allocation}

Under elastic scheduling, system performance is governed by how compute chiplets are dynamically partitioned between prefilling and decoding. We model this behavior as a continuous-time Markov chain (CTMC)~\cite{bolch2006queueing}, where each state represents a particular allocation of chiplets across execution phases.
This formulation captures the dominant system dynamics by modeling chiplet allocation as the primary state variable.

Let $N$ be the total number of compute chiplets. 
A system state is defined by $s \in \{0, \dots, N\}$, 
representing the number of chiplets assigned to prefill, while the remaining $N - s$ chiplets serve decode.
With elastic scheduling, chiplets can be reassigned across phases when demand exceeds the preferred allocation. When prefilling demand exceeds available prefill-oriented chiplets, decoding chiplets are temporarily reassigned, and vice versa. This allows the system to adapt to dynamic workloads. 
This behavior directly reflects the elastic scheduling and dynamic batching mechanisms described in Section~\ref{Sec:Opt}.

\vspace{-3mm}
\subsection{State-Dependent Service Capacity}

For each state $s$, we estimate the system's aggregate processing capacity based on the chiplet types and their assigned roles.
Recall that there are prefill- and decode-optimized chiplets.
Let $c_{p,P}$ and $g_{p,P}$ denote the number active \textit{prefill-optimized chiplets} assigned to \textit{prefill tasks} and their \textit{prefill} token processing rates. 
Similarly, let $c_{d,D}$ and $g_{d,D}$ denote the number active \textit{decode-optimized chiplets} assigned to \textit{decode tasks} and their \textit{decode} token processing rates. 
Elastic scheduling allows \textit{HYDRA} to use decode-optimized chiplets for prefill tasks and vice versa. 
We denote the number of decode-optimized chiplets assigned to prefill tasks as $c_{p,D}$ and 
the corresponding processing rate as $g_{p,D}$. 
Similarly, the number and processing rate of prefill-optimized chiplets assigned to decode tasks are denoted by $c_{d,P}$ and $g_{d,P}$. 
Using these definitions, the aggregate prefilling and decoding capacities can be expressed as:
\begin{align}
G_p(s) &= c_{p,P} \cdot g_{p,P} + c_{p,D} \cdot g_{p,D}  \\
G_d(s) &= c_{d,D} \cdot g_{d,D} + c_{d,P} \cdot g_{d,P} 
\end{align}
These rates depend on both hardware characteristics and workload properties (e.g., sequence length). Thus, they capture both compute and memory constraints, as well as communication overheads.
Since steady-state token generation is limited by decoding, $G_d(s)$ denotes the system throughput in state $s$.

\vspace{-3mm}
\subsection{Throughput and TTFT Estimation}

Computing the steady-state distribution of the CTMC analytically is intractable due to the complex interaction among request arrivals, dynamic batching, and elastic scheduling. Instead, \textit{HYDRA} employs a lightweight fluid-queue model~\cite{bauerle2002optimal} to approximate system dynamics over time. 
The fluid model tracks the aggregate amount of outstanding prefilling and decoding work and updates the chiplet allocation accordingly. 
As workloads are executed, the model tracks the fraction of time spent in each state. Let $\pi_s$ denote the estimated fraction of time spent in state $s$. This approximation captures macroscopic system behavior without requiring fine-grained event-driven simulation. 
Using the prefill/decode processing rates and steady-state probabilities, 
the expected system throughput is:
\begin{equation}
T_p = \sum_{s=0}^{N} \pi_{s} \cdot G_d(s)
\end{equation}
This formulation weights each state's decoding capacity by its occupancy, capturing how workload dynamics and chiplet allocation jointly determine throughput. TTFT is estimated from the same state trajectory, capturing the fundamental throughput--TTFT tradeoff: allocating more chiplets to decoding improves throughput, while allocating more chiplets to prefilling reduces TTFT.

\vspace{-3mm}
\subsection{Generalization to Hybrid Models}
The proposed modeling approach naturally extends to hybrid LLMs with multiple operator types (e.g., Mamba and Attention). In these cases, each state shows how chiplets are assigned to different task classes, and the service rates depend on the characteristics of each operator.
In summary, the proposed estimator captures the main runtime dynamics of hybrid LLM serving and remains computationally efficient while accurately capturing dominant runtime behavior, enabling rapid pruning of suboptimal configurations.

\section{Experiments}\label{sec:eval}
\subsection{Experimental Setup}\label{sec:setup}

\noindent{\textbf{Input models:}}
Evaluations are performed on three representative LLM models: (1) \textit{Nemotron-H-4B}~\cite{blakeman2025nemotron}, a hybrid Transformer–Mamba model that combines Attention and SSM modules; (2) \textit{LLaMA3-7B}~\cite{grattafiori2024llama}, a pure Transformer; (3) \textit{Mamba-2.8B}~\cite{gu2024mamba}, an SSM-based model. These models cover hybrid, Transformer-only, and SSM-only workloads.

\noindent{\textbf{Input datasets:}}
We use four datasets that capture diverse inference patterns: (1) ArXiv-4K~\cite{cohan-etal-2018-discourse} (ArXiv): long scientific documents, representing long prefill/short decode; (2) Bilingual Web Books~\cite{jiang-etal-2023-discourse} (BWB): multilingual narrative text, representing long prefill/long decode; (3) LongWriter-6K~\cite{bailongwriter} (LW): extended-context writing tasks, representing short prefill/long decode; and (4) LMSYS-Chat-1M~\cite{zheng2024lmsys} (Chat): diverse chat interactions, representing short prefill/short decode.
This set spans a broad range of prefill–decode ratios and serving behaviors.

\noindent{\textbf{NoI Specification:}}
We experiment on a passive silicon interposer with a 2D mesh topology, a widely adopted design for large systems~\cite{simba2019}. Die-to-die (D2D) links follow NVIDIA GRS~\cite{GRS2018}, where each 8-lane, 25-Gbps PHY occupies 0.3876~mm$^2$ in 16 nm. Micro-bump area, bandwidth, and latency follow UCIe x64 advanced-package assumptions~\cite{sharma2022universal}. The evaluated D2D bandwidth options are specified in Table~\ref{tab:dse_param}.

\begin{figure*}[t!]
\centering
    \centering
    \includegraphics[width=1.0\linewidth]{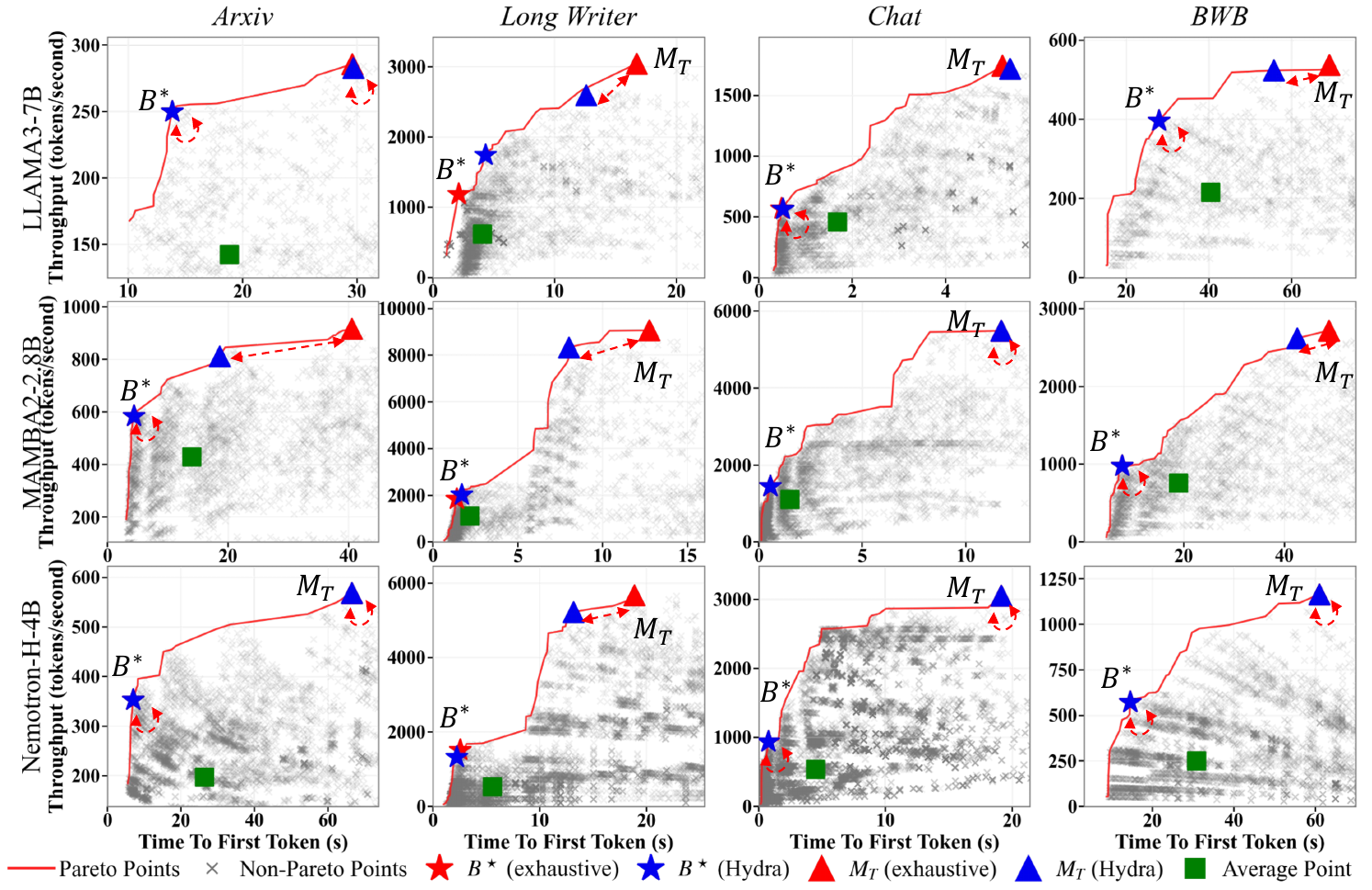}
    \vspace{-6mm}
    \caption{Overview of the DSE on 12 workloads (model-dataset pairs). The red points are the Pareto configurations derived through the exhaustive search of macro-architectures. The blue points are the Pareto configurations estimated by the Markov-based method. $B^{\star}$: Configurations achieving the highest throughput-TTFT ratio; $M_T$: Configurations achieving the best throughput.}
    \label{fig:DSE_overview}
\vspace{-6mm}
\end{figure*}

\begin{table}[t]
\centering
\caption{Design space exploration parameters. Some configurations are shared across all models.}
\label{tab:dse_param}
\setlength{\tabcolsep}{4pt} 

\begin{tabular}{lccc}
\toprule
\textbf{Chiplet Counts} & \textbf{LLAMA3} & \textbf{MAMBA2} & \textbf{Nemotron-H} \\
\midrule
Total & \multicolumn{3}{c}{\hspace{-3.5mm}$24$} \\
$N_{Ap}$ (Prefill) & $1$ -- $21$ & --          & $1$ -- $19$ \\
$N_{Ad}$ (Decode)  & $1$ -- $21$ & --          & $1$ -- $19$ \\
$N_{Mp}$ (Prefill) & --          & $1$ -- $21$ & $1$ -- $19$ \\
$N_{Md}$ (Decode)  & --          & $1$ -- $21$ & $1$ -- $19$ \\
$N_{HBM3}$ & \multicolumn{3}{c}{\hspace{-3.5mm}$2$ -- $16$} \\

\midrule
\multicolumn{4}{l}{\textbf{Other Configs}} \\
D2D BW     & \multicolumn{3}{c}{$256,384,512,640$ GB/s} \\
Max Batch Size & \multicolumn{3}{c}{$2,4,8,16,32,64$} \\
\bottomrule
\end{tabular}
\vspace{-3mm}
\end{table}

\noindent{\textbf{Interposer/die area constraints:}}
We assume 2700–3000~mm$^2$ interposer area, comparable to state-of-the-art packages such as NVIDIA B200~\cite{nvidia_blackwell_datasheet_2024} and AMD MI350~\cite{amd_mi350_series_2025}. This range is comparable to state-of-the-art accelerator packages and permits large-scale heterogeneous chiplet systems. Individual chiplet area is capped at approximately 121~mm$^2$, comparable to the footprint of an HBM stack, to simplify floorplanning and avoid the steep cost increase associated with large dies, following standard chiplet cost models~\cite{Feng_2022}.

\noindent{\textbf{Chiplet library:}}
The chiplet library comprises four LLM accelerator chiplets: MARCA~\cite{li2024marca} optimized for the prefill and decode phases of Mamba ($M_p$ and $M_d$), and TSTC~\cite{liu2023tstc} optimized for the prefill and decode phases of Transformer ($A_p$ and $A_d$). Prefill-oriented chiplets allocate area approximately evenly between compute and SRAM, whereas decode-oriented chiplets allocate roughly twice as much area to memory resources as to computation to better support bandwidth-intensive decoding. 

As discussed in Section~\ref{sec:systemspec}, higher NoI bandwidth requires additional communication resources on the chiplet, reducing the area available for compute units and on-chip SRAM. In addition, the system employs HBM3 chiplets for model parameters, KV caches, and intermediate states. To ensure fair comparison across heterogeneous designs, all components are normalized to a 22~nm CMOS node using DeepScale~\cite{stillmaker2017scaling}. This choice enables comparing designs originally proposed for different process nodes using a unified cost and area model.

\noindent{\textbf{LLM inference evaluation:}} 
LLM inference is evaluated using an in-house event-driven simulator that models computation, memory behavior, and NoI communication at chiplet granularity. It tracks per-chiplet task queues, execution progress, and link utilization, enabling accurate modeling of contention and overlap across computation and communication. 
Memory chiplet performance is characterized using Ramulator~\cite{luo2023ramulator}, with capacity and bandwidth parameters derived from vendor-level specifications~\cite{park2022192}. On-chip SRAM buffer areas are obtained from CACTI~\cite{balasubramonian2017cacti}, while special function units (SFUs) for non-linear operators are modeled using FlexSFU~\cite{andri2025flex}. The performance of MARCA and TSTC kernels is modeled using the microarchitectural parameters and kernel-level characterizations reported in prior work, accounting for operation type, data dependencies, and available parallelism.

All simulations are conducted on an AMD Ryzen Threadripper 7985WX CPU~\cite{amd_threadripper_aec}. Each configuration is simulated for 100 seconds of workload time.

\vspace{-2mm}
\subsection{The DSE on Macro-architectures}\label{sec:dse_res}

The goal of this section is to analyze how chiplet composition, NoI bandwidth provisioning, and batch sizing shape the throughput–TTFT trade-off when dynamic optimizations are disabled. 
Hence, we explore macro-architectural configurations under \textit{static request batching} and 
\textit{static task scheduling}~
\cite{adiletta2025democratizing} to expose fundamental architectural trade-offs, using the design parameters summarized in Table~\ref{tab:dse_param}. 

\noindent\textbf{Methodology:} To efficiently explore this design space, we first apply a fast performance estimation stage to identify candidate Pareto-optimal configurations in the throughput \rev{(TP)}–TTFT space. This stage uses our Markov–based estimator (Section~\ref{sec:markov}) to approximate system performance,
reducing the number of simulated design points by 2520–8520$\times$ relative to exhaustive exploration. The proposed Markov-based estimator is sufficiently general to model both dynamic and static execution. In this experiment, static batching and scheduling correspond to a special case in which chiplet allocation remains fixed and runtime reassignment is disabled. For validation, we compare the predicted Pareto points against exhaustive event-driven simulation that enumerates all feasible configurations under the same constraints.

\noindent\textbf{Design trade-off:} Fig.~\ref{fig:DSE_overview} summarizes the macro-architecture design space exploration results across all model–dataset pairs. Each subplot shows the throughput and TTFT of all simulated configurations. The red curve traces the Pareto frontier obtained from exhaustive simulation, while the gray points correspond to dominated configurations. Across all workloads, configurations that increase throughput typically allocate more resources to decoding, either through larger effective batch sizes or higher NoI bandwidth. While this improves steady-state token generation, it often increases TTFT by reducing resources available to prefilling and increasing resource contention.  This trade-off fundamentally arises between prefill latency and decode throughput under shared system resources, and highlights the need to select appropriate operating points rather than optimizing a single metric.

From the Pareto frontier, we select two representative configurations for further analysis. The configuration labeled \textit{$B^{\star}$ is a balanced point} that maximizes the throughput to TTFT ratio among configurations whose throughput exceeds the design-space average. The configuration labeled \textit{$M_T$ maximizes throughput} and represents an aggressively provisioned, throughput-oriented design. As our study targets high-throughput multi-tenant serving systems, subsequent evaluations focus on these two operating points, which reflect the two design objectives considered in this study.

\noindent\textbf{Accuracy of the Markov–based performance estimator:} 
The blue markers in Fig.~\ref{fig:DSE_overview} (\textcolor{blue}{$\star$}, \textcolor{blue}{$\blacktriangle$}) indicate the Pareto-optimal configurations predicted by the proposed Markov-based estimator, while the red markers (\textcolor{red}{$\star$}, \textcolor{red}{$\blacktriangle$}) show the corresponding simulation results. 
For most workloads, the estimator closely tracks the simulated Pareto frontier, with only modest discrepancies. For example, in the Nemotron-H–LW workload, the balanced configuration $B^{\star}$ exhibits a 8\% throughput gap, while the throughput of $M_T$ deviates by approximately 12\% under static scheduling. These gaps arise because the estimator abstracts away fine-grained queueing effects, transient resource contention, and scheduling decisions, whereas the simulator explicitly models these interactions. 
Nevertheless, the estimator preserves the relative ordering of configurations and accurately identifies high-quality Pareto candidates, making it effective for pruning inferior designs prior to detailed simulation.

\vspace{-2mm}

\begin{figure*}[t!]
\centering
    \centering
    \includegraphics[width=1\linewidth]{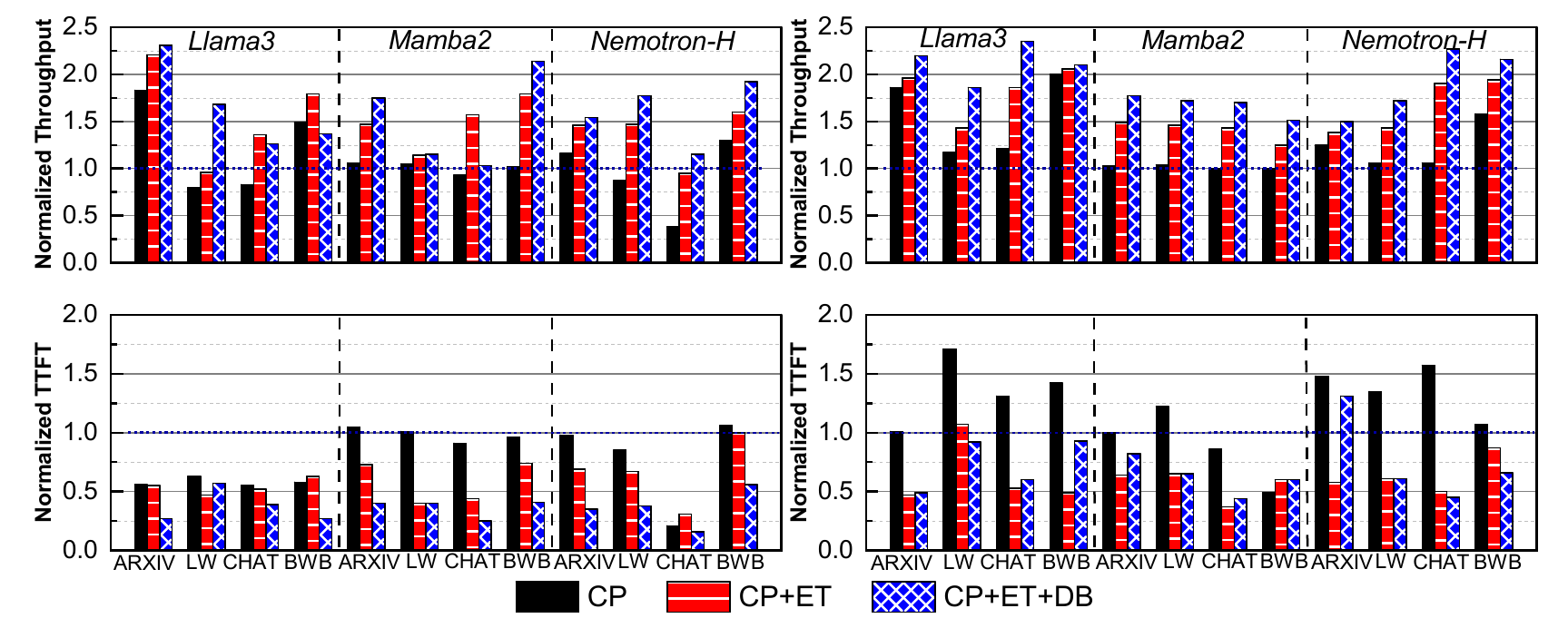}
    \caption{Ablation study of communication-aware placement (CP, Section~\ref{sec:chip_placmt}), elastic task scheduling (ET, Section~\ref{sec:task_sched}), and dynamic batching (DB, Section~\ref{sec:dynamic_batch}) strategies. It reports results for 12 model-dataset pairs on the x-axis for the $B^{\star}$ and $M_T$.}
    \label{fig:Hydra_ablation}
\vspace{-6mm}
\end{figure*}

\subsection{\rev{Specialization--Generality Trade-Off Study}}\label{sec:tradeoff}


\definecolor{MostSpecColor}{RGB}{235,245,255}
\definecolor{MedSpecColor}{RGB}{245,245,245}
\definecolor{GeneralColor}{RGB}{255,245,230}

\newcommand{\tabhdr}[1]{{\fontsize{5.0}{5.0}\selectfont\textbf{#1}}}

\begin{table}[b!]
\centering
\caption{\rev{Specialization--generality tradeoff across hybrid LLMs. Each row is the DSE target model set, and each column is the evaluation workload. Values are normalized to the workload-specific single-model optimum. (a) reports $B^{\star}$ with normalized TP/TTFT; (b) reports $M_T$ with normalized TP.}}
\label{tab:specialization_tradeoff}

\setlength{\tabcolsep}{3.2pt}
\renewcommand{\arraystretch}{0.85}

{\footnotesize\textbf{(a) $B^{\star}$: norm. TP/TTFT score $\uparrow$}}\\[0.25em]
\resizebox{\linewidth}{!}{
{\fontsize{5.0}{5.0}\selectfont
\begin{tabular}{lccc}
\toprule
\tabhdr{DSE Target} & \tabhdr{Jamba} & \tabhdr{Zamba} & \tabhdr{Nemo} \\
\midrule
\rowcolor{MostSpecColor}
Jamba & 1.00 & \textcolor{red}{0.42} & \textcolor{red}{0.31} \\
\rowcolor{MostSpecColor}
Zamba & \textcolor{red}{0.91} & 1.00 & \textcolor{red}{0.76} \\
\rowcolor{MostSpecColor}
Nemo & \textcolor{red}{0.87} & \textcolor{red}{0.32} & 1.00 \\
\rowcolor{MedSpecColor}
Jamba + Zamba & 0.97 & 0.94 & \textcolor{red}{0.71} \\
\rowcolor{MedSpecColor}
Jamba + Nemo & 0.91 & \textcolor{red}{0.29} & 0.92 \\
\rowcolor{MedSpecColor}
Zamba + Nemo & \textcolor{red}{0.96} & 0.91 & 0.86 \\
\rowcolor{GeneralColor}
Jamba + Zamba + Nemo & 0.96 & 0.91 & 0.86 \\
\bottomrule
\end{tabular}
}
}

\vspace{0.45em}

{\footnotesize\textbf{(b) $M_T$: norm. TP $\uparrow$}}\\[0.25em]
\resizebox{\linewidth}{!}{
{\fontsize{5.0}{5.0}\selectfont
\begin{tabular}{lccc}
\toprule
\tabhdr{DSE Target} & \tabhdr{Jamba} & \tabhdr{Zamba} & \tabhdr{Nemo} \\
\midrule
\rowcolor{MostSpecColor}
Jamba & 1.00 & \textcolor{red}{0.44} & \textcolor{red}{0.36} \\
\rowcolor{MostSpecColor}
Zamba & \textcolor{red}{0.97} & 1.00 & \textcolor{red}{0.80} \\
\rowcolor{MostSpecColor}
Nemo & \textcolor{red}{0.98} & \textcolor{red}{0.62} & 1.00 \\
\rowcolor{MedSpecColor}
Jamba + Zamba & 0.97 & 1.00 & \textcolor{red}{0.80} \\
\rowcolor{MedSpecColor}
Jamba + Nemo & 0.98 & \textcolor{red}{0.62} & 1.00 \\
\rowcolor{MedSpecColor}
Zamba + Nemo & \textcolor{red}{0.99} & 0.96 & 0.84 \\
\rowcolor{GeneralColor}
Jamba + Zamba + Nemo & 0.99 & 0.96 & 0.84 \\
\bottomrule
\end{tabular}
}
}

\vspace{0.4em}
{\footnotesize
\begin{tabular}{@{}ll@{\hspace{1.0em}}ll@{}}
\textcolor{MostSpecColor}{\rule{1.2em}{0.75em}} & Most-specialized &
\textcolor{MedSpecColor}{\rule{1.2em}{0.75em}} & Medium-specialized \\
\textcolor{GeneralColor}{\rule{1.2em}{0.75em}} & General &
\textcolor{red}{red} & off-target evaluation \\
\end{tabular}
}
\end{table}

\rev{
We further evaluate how well configurations optimized for one hybrid LLM transfer to other hybrid LLMs, exposing the trade-off between model specialization and cross-model generality. This study uses the Chat dataset and three representative hybrid LLM workloads: Jamba-tiny~\cite{lenz2025jamba} (Jamba), Zamba2-7B~\cite{glorioso2024zamba} (Zamba), and Nemotron-H-4B~\cite{blakeman2025nemotron} (Nemo). 
Jamba contains 2 Attention and 14 Mamba blocks, Zamba contains 13 Attention and 81 Mamba blocks, and Nemotron-H contains 4 Attention and 24 Mamba blocks.
For each DSE target, \textit{HYDRA} explores the same macro-architectural design space as in Section~\ref{sec:dse_res}. We consider three specialization levels: (i) single-model DSE, which produces the most specialized configuration, (ii) two-model DSE, which produces a partially generalized configuration, and (iii) three-model DSE, which produces a general configuration optimized across all workloads. The corresponding results are shown as blue, gray, and orange rows in Table~\ref{tab:specialization_tradeoff}, respectively.
}

\rev{
Table~\ref{tab:specialization_tradeoff} reports the resulting specialization--generality tradeoff for the two operating objectives in Section~\ref{sec:dse_res}: the balanced point ($B^\star$) and the maximum-throughput point ($M_T$). Values are normalized to the best result for each workload, so 1.00 denotes the workload-specific optimum. Red entries indicate off-target evaluations, where a configuration is tested on a model excluded from its DSE target.
}

\rev{
First, the $B^\star$ results in Table~\ref{tab:specialization_tradeoff}(a) show that single-model specialization gives the best result on the target workload, but can transfer poorly to other hybrid models. For example, the Jamba-specialized design achieves 1.00 on Jamba, but drops to 0.42 on Zamba and 0.31 on Nemo. This demonstrates that balanced configurations can over-specialize to a particular layer composition and serving behavior. In contrast, mixture-aware DSE improves the weakest transfer cases. The all-model configuration raises Nemo from 0.31 to 0.86 compared with the Jamba-specialized design ($2.77\times$), and raises Zamba from 0.32 to 0.91 compared with the Nemo-specialized design ($2.84\times$). The all-model configuration limits the worst-case degradation to 14\% relative to each workload's specialized optimum.
}

\rev{
Second, the $M_T$ results in Table~\ref{tab:specialization_tradeoff}(b) show a similar trend, although throughput-oriented designs transfer more effectively because the objective is dominated by sustained decode capacity and batching efficiency. For example, the Zamba-specialized design preserves 97\% of Jamba's maximum throughput and 80\% of Nemo's maximum throughput. However, the Jamba-specialized design remains poorly matched to Zamba and Nemo, reaching only 0.44 and 0.36, respectively, reflecting architectural choices that are tuned to Jamba's workload characteristics. By comparison, the all-model $M_T$ configuration improves these weak transfer cases to 0.96 and 0.84 ($2.18\times$ and $2.33\times$), while reducing Jamba throughput by only 1\%. Similarly, it improves Zamba from 0.62 to 0.96 compared with the Nemo-specialized design, at the cost of reducing Nemo from 1.00 to 0.84.
}

\rev{
Overall, these results highlight the deployment tradeoff. For services dedicated to a single stable model, specialized configurations provide the best performance. For platforms expected to support multiple or evolving hybrid LLMs, mixture-aware DSE offers a more robust configuration by substantially improving worst-case performance on unseen models while remaining close to each model's specialized optimum.
}

\vspace{-2mm}
\subsection{Evaluation of Placement Strategies}\label{sec:placmt}

We next evaluate the impact of chiplet placement strategies on system performance. While the macro-architectural configuration determines available resources, placement governs how efficiently these resources are interconnected. Fig.~\ref{fig:Hydra_ablation} first reports the effect of communication-aware placement, and the next subsection (Section~\ref{sec:scheduling}) uses Fig.~\ref{fig:Hydra_ablation} to further evaluate elastic scheduling and dynamic batching for $B^\star$ and $M_T$ across all 12 workloads. Results are normalized to a round-robin placement with static scheduling used in Section~\ref{sec:dse_res}. 

The black bars correspond to our communication-aware placement (CP) strategy, which clusters chiplets based on their dominant communication patterns. 
Across all workloads, CP improves throughput by 3\%--100\% while maintaining comparable or lower TTFT for $M_T$. 
For $B^{\star}$, CP consistently achieves a higher throughput-to-TTFT ratio than the baseline. In several cases, CP simultaneously improves both metrics by enabling larger effective batch sizes. For example, in the Nemotron-H-ArXiv workload under $B^{\star}$, CP improves throughput by $1.3\times$ with only a $1.06\times$ increase in TTFT.

\begin{figure}[t]
\centering
    \centering
    \includegraphics[width=1\linewidth]{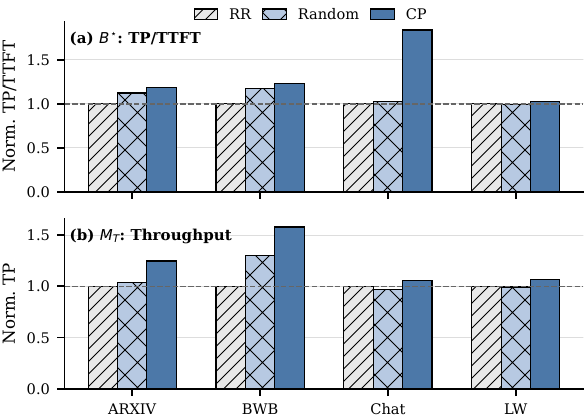}
    \caption{\rev{Comparison of chiplet placement strategies across Nemotron-H workloads.}}
    \label{fig:placer_baselines}
\vspace{-6mm}
\end{figure}


\rev{
To further show the impact of placement, we add a random placement baseline evaluated by the same Nemotron-H workloads and hardware configurations as in Section~\ref{sec:dse_res}. All placement baselines keep memory chiplets on the outer ring of the interposer. The round-robin (RR) baseline distributes accelerator chiplets and model weights in round-robin order, while the random baseline randomly places accelerator chiplets and randomly assigns model weights to HBM chiplets, \revcam{requiring negligible design-time optimization overhead.}
}


\rev{
Fig.~\ref{fig:placer_baselines} reports the resulting TP/TTFT for $B^{\star}$ and throughput for $M_T$ on the ARXIV, BWB, Chat, and LongWriter datasets. Random placement behaves similarly to RR, indicating that simply distributing chiplets across the interposer does not consistently improve communication locality. On average, CP improves TP/TTFT by $1.29\times$ and throughput by $1.22\times$ relative to RR. Compared with random placement, CP improves TP/TTFT by $1.04\times$--$1.80\times$ and throughput by $1.07\times$--$1.22\times$. These results show that robust placement gains require explicitly aligning compute chiplets with their dominant HBM communication partners rather than relying on arbitrary chiplet distributions.
}

\begin{figure}[t]
\centering
    \centering
    \includegraphics[width=1\linewidth]{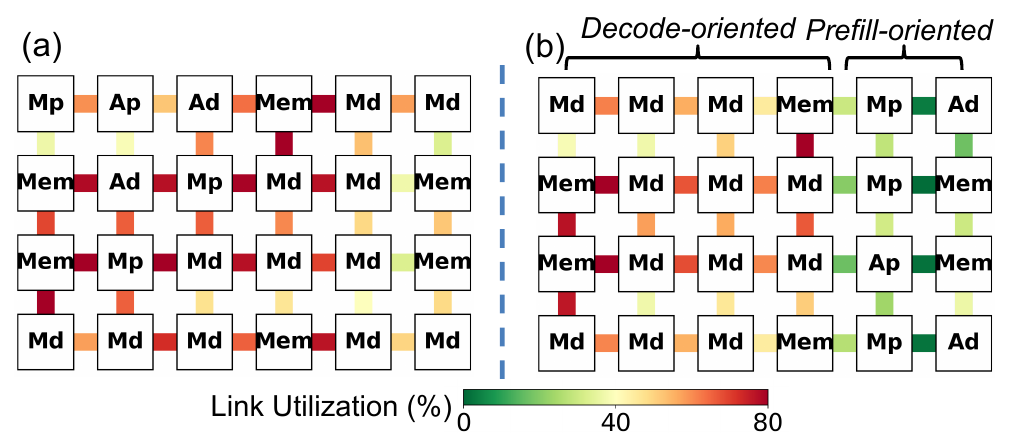}
    \caption{NoI heatmaps of bandwidth utilization that compare the round-robin (a) and communication-aware placement (b).}
    \label{fig:bwcomp}
\vspace{-4mm}
\end{figure}

To understand the source of these improvements, Fig.~\ref{fig:bwcomp} visualizes NoI link-utilization heatmaps for the Nemotron-H–LW workload under two placement strategies. The round-robin placement (left) exhibits widespread D2D bandwidth pressure, indicating frequent long-distance communication between HBM and compute chiplets. In contrast, CP (right) significantly reduces bandwidth pressure, as reflected by lighter-colored links, by minimizing average hop counts and localizing data movement.
\rev{Although CP reduces average hop count and removes widespread NoI pressure, it cannot eliminate hotspots created by the intrinsic bandwidth demand of steady-state decoding. The remaining high-utilization links are therefore expected under fixed NoI bandwidth.}
In addition, CP spatially separates prefilling-oriented and decoding-oriented chiplets, which reduces cross-phase interference on shared NoI resources. While this separation can slightly increase TTFT in some cases due to reduced opportunistic sharing, it substantially improves throughput by mitigating sustained NoI congestion during steady-state decoding. 

This observation demonstrates that reducing communication distance is more effective than simply increasing compute parallelism, highlighting that communication locality, rather than raw compute provisioning, is often the dominant bottleneck in chiplet-based LLM serving. This aligns with \textit{HYDRA}’s design philosophy of co-optimizing architecture and data movement. 

Together, these results show that placement establishes baseline system behavior but cannot adapt to dynamic arrivals or transient resource imbalance. The next subsection evaluates how elastic scheduling and dynamic batching address them.

\vspace{-2mm}
\subsection{Evaluation of Elastic Scheduling and Dynamic Batching}\label{sec:scheduling}

Fig.~\ref{fig:Hydra_ablation} also evaluates the impact of runtime elastic task scheduling and dynamic batching on the two representative configurations, $B^{\star}$ and $M_T$, across all 12 workloads. The red horizontally hatched bars enable runtime elastic task scheduling on top of static batching, while the blue crosshatched bars further incorporate dynamic request batching.

\noindent\textbf{Elastic scheduling} improves performance across all workloads by dynamically redistributing work across chiplets when runtime imbalance arises. Under static mapping, task assignments are fixed offline, and transient effects, such as prolonged decoding phases or bursty request arrivals, can create pipeline stalls and leave resources underutilized. Elastic scheduling mitigates this by migrating ready tasks to underutilized chiplets, guided by task–accelerator affinity and bounded communication overhead. This enables adaptive load balancing, leading to 3\%--150\% higher throughput and 1.02$\times$--3.14$\times$ lower TTFT.

\begin{figure}[t]
\centering
    \centering
    \includegraphics[width=0.9\linewidth]{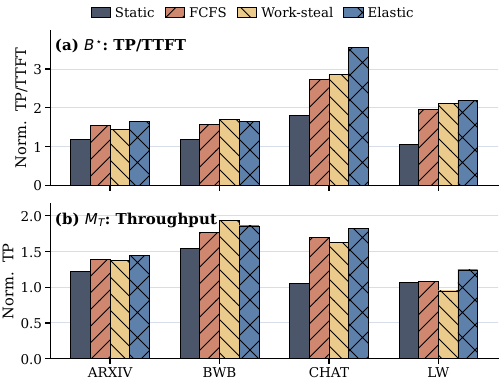}
    \caption{\rev{Comparison of task schedulers across Nemotron-H workloads.}}
    \label{fig:scheduler_baselines}
\vspace{-5mm}
\end{figure}


\rev{
To further evaluate the effect of task scheduling, we compare four task-mapping policies using the same Nemotron-H workloads and hardware configurations as in Section~\ref{sec:dse_res}. 
For each workload, only the runtime task scheduler is changed. Static mapping always follows the placement-selected chiplet assignment. The First-Come-First-Served (FCFS) baseline~\cite{silberschatz2018osc} assigns each incoming task to a compatible chiplet in round-robin order, ignoring placement preference and locality. Work-stealing~\cite{chen2018architectural} dynamically migrates queued tasks from heavily loaded chiplets to lightly loaded compatible chiplets to improve load balance. In contrast, \textit{HYDRA}'s elastic scheduler preserves the placement-preferred chiplet when the imbalance is mild, but selectively reassigns tasks based on both queue occupancy and communication distance to the associated memory chiplets. This design retains communication locality while still exploiting dynamic load balancing opportunities.
}


\rev{
Fig.~\ref{fig:scheduler_baselines} reports the resulting TP/TTFT for $B^{\star}$ and throughput for $M_T$ across the ARXIV, BWB, Chat, and LongWriter datasets. Compared with static mapping, elastic scheduling improves TP/TTFT by $1.40\times$--$2.04\times$ and improves maximum throughput by $1.17\times$--$1.73\times$. Compared with FCFS, elastic scheduling improves the target metric by $1.04\times$--$1.30\times$ in all cases. Compared with work-stealing, elastic scheduling achieves the best result in six of the eight cases, improving the target metric by up to $1.31\times$. In the two remaining BWB cases, elastic scheduling remains within 4\% of work-stealing. This is because work-stealing aggressively reacts to queue imbalance, but it is not communication-aware. Overall, these results show that elastic scheduling provides most of the load-balancing benefit of aggressive dynamic reassignment while considering communication locality, unlike FCFS and work-stealing.
}

\noindent\textbf{Dynamic batching} further amplifies these gains by increasing the effective batch sizes at runtime. By decoupling memory allocation from admission through virtualized memory blocks, \textit{HYDRA} can admit new requests as long as aggregate compute and memory budgets permit. The batching controller continuously monitors token-level progress and adjusts the number of in-flight requests to maximize utilization while bounding TTFT increase. 
Dynamic batching improves memory utilization and throughput of attention-based models, whose KV-cache footprint grows with sequence length. Though SSM-based models exhibit relatively fixed memory footprints determined by state dimensions, they still benefit from continuous request batching. Across all workloads, dynamic batching achieves a 15\% throughput gain while reducing TTFT by 23\% on average.

\begin{figure}[t]
\centering
    \centering
    \includegraphics[width=1\linewidth]{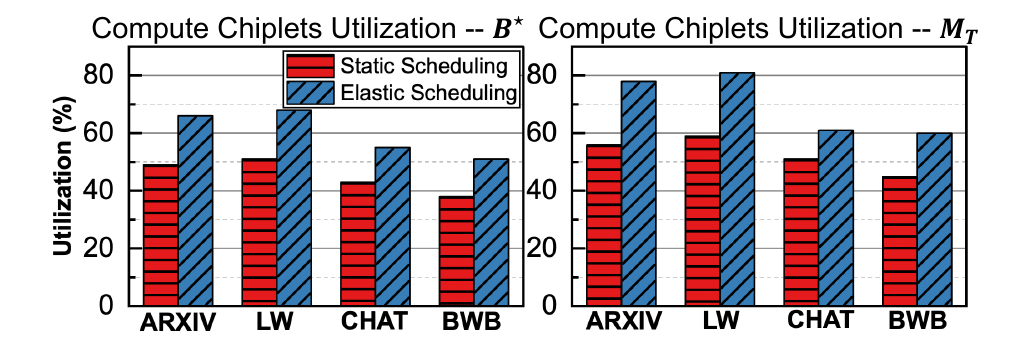}
    \caption{Comparison of compute-chiplet utilization across Nemotron-H workloads.}
    \label{fig:utilcomp}
\vspace{-5mm}
\end{figure}

\noindent\textbf{Quantitative summary:} Elastic scheduling primarily improves load balance, while dynamic batching improves timely resource allocation. 
This combination achieves 1.55$\times$ the throughput and 43.7\% lower TTFT on average relative to the static baseline, with throughput gains reaching up to 2.3$\times$. These improvements correlate with higher and more balanced compute-chiplet utilization, as shown in Fig.~\ref{fig:utilcomp} for Nemotron-H across four datasets. 
These results demonstrate that runtime adaptivity is essential for fully exploiting optimized macro-architectural configurations. While placement establishes efficient communication structure, dynamic scheduling and batching are required to handle workload variability and achieve sustained high utilization in multi-tenant LLM serving.

\vspace{-3mm}
\subsection{Performance Model for Fast DSE}\label{sec:fastdse}

Table~\ref{tab:exploration_runtime} compares the efficiency and effectiveness of different design-space exploration strategies. Exhaustive simulation evaluates all feasible configurations under dynamic arrivals, batching, and scheduling, but incurs prohibitively long runtimes, ranging from \textit{multiple days to over a week}, due to detailed NoI communication, memory management, and runtime contention models. In contrast, the roofline-based approach completes within minutes by relying on static kernel bounds and idealized execution assumptions, but its limited modeling fidelity leads to suboptimal design choices. 
The best configuration it finds achieves only 74\% of the achievable throughput.

\textit{HYDRA}’s Markov-based estimator incorporates workload characteristics and runtime behavior, including elastic scheduling and dynamic batching, to identify high-quality configurations in 4–15 minutes, comparable to the roofline baseline. Across all workloads, \textit{HYDRA} recovers 93\% of the maximum throughput found by exhaustive simulation on average. Moreover, the Pareto-optimal set closely tracks the exhaustive Pareto frontier and contains optimal configurations in some cases. These results demonstrate that the Markov-based estimator accurately captures the dominant runtime dynamics that govern multi-tenant LLM serving performance. This enables efficient search while supporting fast, accurate, and scalable exploration of system-level trade-offs in heterogeneous chiplet architectures for LLM serving.








\begin{table}[ht]
\centering
\caption{Average exploration time for different search strategies and maximum-throughput recovery. Each exploration covers four datasets per model.}
\label{tab:exploration_runtime}

\resizebox{0.9\columnwidth}{!}{%
\begin{tabular}{l ccc}
\toprule
\textbf{Model} & \makecell{\textbf{Exhaustive}} & \makecell{\textbf{Roofline~\cite{williams2009roofline}}} & \makecell{\textbf{HYDRA}} \\
\midrule

\textbf{LLAMA3-7B}    & 2 d. 8 h. & 4 min & 4 min \\ 
\textbf{MAMBA2-2.8B}  & 3 d. 1 h.& 4 min & 4 min \\ 
\textbf{Nemotron-H-4B}& 8 d. 8 h.& 15 min & 15 min \\ 
\midrule
\textbf{Max Throughput}& 100\% & 74\% & 93\% \\

\bottomrule
\end{tabular}%
}
\vspace{-5mm}
\end{table}

\section{Conclusion}\label{sec:conclusion}

This paper presented \textit{HYDRA}, a design space exploration framework for heterogeneous chiplet-based systems targeting dynamic hybrid LLM serving. \textit{HYDRA} jointly models chiplet placement, request batching, runtime scheduling, and fast performance estimation to capture the interaction between architecture and workload dynamics. Across hybrid, Mamba, and Transformer workloads, the proposed optimization strategies \revcam{deliver 1.55$\times$ the throughput and 43.7\% lower TTFT on average, with throughput gains reaching up to 2.3$\times$}, and the Markov-based estimator enables efficient exploration that reduces DSE time from days to minutes. These results show that effective hybrid LLM serving requires co-optimizing compute, memory, communication, and runtime control policies. Overall, \textit{HYDRA} provides a practical and scalable framework for exploring heterogeneous chiplet architectures for next-generation multi-tenant LLM systems.

\rev{
This work focuses on throughput and TTFT. Power, thermal, and reliability remain important design considerations, particularly as chiplet counts and communication bandwidth continue to increase. Since HYDRA already models placement, communication, memory utilization, and runtime scheduling, it can be extended with power and thermal models for corresponding DSE in the future work.
}

\vspace{1mm}
\noindent \textit{Disclosure}: Dr. Ogras is affiliated with Samsung Austin Research \& Development Center and Advanced Computing Lab (SARC/ACL). This relationship has been approved under applicable outside activities policies.

AI tools are used to improve the clarity and readability portions of the manuscript. All technical content, results, and conclusions were verified by the authors.

\bibliographystyle{IEEEtran}
\bibliography{References/paper}

\ifthenelse{\boolean{RevResponse}}{
\clearpage
\section*{Responses to Reviewer Comments}
We thank the reviewers for their constructive feedback on our manuscript titled "HYDRA: A Heterogeneous Chiplet DSE Framework for Serving Dynamic Hybrid LLM Workloads".
All changes in the revised manuscript are highlighted in \rev{blue typeface}.
The major changes in the revised manuscript are:
\begin{itemize}[leftmargin=*]
    \item \textbf{Expanded cross-model evaluation:} We added new experiments to study the specialization--generality tradeoff of HYDRA-generated hardware configurations, and their robustness to unseen hybrid LLMs with different Attention-to-Mamba ratios.

    \item \textbf{Additional baselines:} We added new placement and scheduling baselines, including random placement, first-come-first-serve (FCFS) scheduling, and work-stealing scheduling.

    \item \textbf{Improved methodology clarity:} We expanded the methodology and evaluation sections to clarify the computation of $t_{\text{comp}}$ and $t_{\text{comm}}$, the task-chiplet affinity used by the scheduler, the need for detailed simulation, and the interpretation of the NoI hotspot results.
\end{itemize}

\vspace{-2mm}
\section*{\textbf{Reviewer \#1}}



\bh{Comment 1.1:} 
\begin{condensed}
The tradeoffs between specialization to a particular model vs a general-purpose design that is good for many models are not clear. The main piece missing from this work is evaluating ``specialized" designs vs more general purpose designs. \textit{HYDRA} enables the discovery of more specialized designs which will give gains for specific models that it is designed for. However, 2 things are unclear. What is the tradeoff between how specialized the design is versus how much efficiency is gained? So, if the system is designed for exactly 1 model, how much gain over general-purpose. If it is designed for 2, 3, 4, etc. models, then how does efficiency get worse?
\end{condensed}

\bh{Response 1.1:}
We agree that the tradeoff between model-specific specialization and cross-model generality is a key consideration for practical deployment. Therefore, we performed new experiments by sweeping the degree of specialization to find the Pareto-optimal designs targeting: (i) a single model, (ii) a mixture of two models, and (iii) a mixture of three models from the hybrid model set (i.e., Nemotron-H, Jamba, and Zamba). For each setting, we report throughput and TTFT relative to the most-specialized single-model design. These results quantify how efficiency changes as the target design becomes more general, helping practitioners choose between model-specific and mixture-aware hardware configurations. 
Indeed, the additional analysis (reproduced below) shows that configurations optimized across multiple models provide significantly better robustness, while incurring only a modest loss relative to the best model-specific design.
The corresponding revisions are included in \textbf{Section~\ref{sec:tradeoff}}, on \textbf{Page 10}, \textbf{Rows 780–839}.

\bh{Comment 1.2:} Second, how resilient are the HYDRA designs for models that have not been seen before? It seems like the big danger of designing specialized systems is that they do not work on a model that comes out a year from now, whereas a general-purpose design would be resilient. Given the rapidly changing landscape of LLMs, the dangers of specialization should be discussed/evaluated.

\bh{Response 1.2:} We agree that resilience to unseen models is important in practice. To quantify the risk of over-specialization, the cross-model study in Response 1.1 evaluates each configuration on hybrid LLMs excluded from its DSE target. Table~\ref{tab:specialization_tradeoff} highlights these cases in red: single-model configurations can fall to 0.31--0.44 normalized performance, whereas the all-model configuration maintains at least 0.83 for both $B^{\star}$ and $M_T$ across Jamba, Zamba, and Nemo. The corresponding revisions are included in \textbf{Section~\ref{sec:tradeoff}}, on \textbf{Page 10}, \textbf{Rows 780–839}.

\bh{Comment 1.3:} Importance of state-space models is not clear.
Given the effectiveness of the estimator, it is unclear if detailed simulations are even necessary in the final system.

\bh{Response 1.3:} \textit{Regarding the importance of SSMs}, we revised the introduction to clarify their role in emerging hybrid LLMs. Unlike self-attention, whose KV-cache storage and memory traffic grow with context length, SSMs maintain a compact recurrent state and provide linear-time sequence processing. These properties improve long-context serving efficiency and have motivated recent hybrid models, including Jamba, Zamba, and Nemotron-H. Their heterogeneous Attention and SSM computations are precisely the workloads targeted by \textit{HYDRA}. \textit{Regarding the estimator}, we clarify that it complements rather than replaces detailed simulation. The estimator rapidly prunes the design space and ranks promising configurations, reducing DSE time from days to minutes. Detailed simulation is then applied only to shortlisted candidates to validate throughput and TTFT while capturing transient contention, NoI and memory effects, dynamic batching, and pipeline imbalance that the estimator abstracts away. Thus, detailed simulation remains part of the offline DSE methodology, not the deployed runtime system. These revisions appear in \textbf{Section~\ref{sec:intro}} on \textbf{Page~1}, \textbf{Rows 31–37}, and \textbf{Page~2}, \textbf{Rows 113–120}.





\bh{Comment 1.4:} Minor comments: In Fig 8, there are some missing red circles/triangles. Where are those? Fig 9 is quite ugly, please use different colors for the bars. In Fig 10, please explain why there are still notable hotspots on the lefthand portion of the system. Why was the communication aware unable to reduce those hotspots?

\bh{Response 1.4:} We handle all these figure-related comments. For \textbf{Fig.~\ref{fig:DSE_overview}}, the missing red markers are due to overlap: in several subplots, Markov-predicted points coincide with exhaustive-search points, so we add small offsets/annotations to make them visible. For \textbf{Fig.~\ref{fig:Hydra_ablation}}, we redraw the bars with clearer colors and distinguishable hatching. For \textbf{Fig.~\ref{fig:bwcomp} (Fig. 10 referenced by the reviewer)}, communication-aware placement minimizes average hop count but cannot eliminate the intrinsic bandwidth demand of decoding under the chosen NoI provisioning. We add this explanation in \textbf{Section~\ref{sec:placmt}} on \textbf{Page~12}, \textbf{Rows 889–893}.



\vspace{-2mm}
\section*{\textbf{Reviewer \#2}}

\bh{Comment 2.1:}
There are many ways in which tasks are scheduled in the literature. The authors should do a study of different schemes - and show which is better. They simply say that they do load balancing in a an adaptive way. How? More details are needed. There are so many schemes in the literature...

\bh{Response 2.1:}
We agree that the original manuscript did not sufficiently explain how \textit{HYDRA}'s elastic scheduler operates or compare it with representative policies. We therefore revised Section~\ref{sec:scheduling} to clarify that \textit{HYDRA} begins with a communication-aware static mapping and selectively reassigns ready tasks to compatible underutilized chiplets based on per-chiplet queue occupancy. Unlike purely load-driven schedulers, it also considers communication distance to preserve locality while improving utilization.

We also added a study comparing four task-mapping policies: static affinity mapping, first-come-first-served (FCFS), work-stealing, and HYDRA's elastic scheduler. To isolate scheduling effects, we use the same Nemotron-H workloads and fix each hardware configuration to the $B^{\star}$ or $M_T$ point selected in Section~\ref{sec:dse_res}. The results show that runtime reassignment improves over static mapping, while HYDRA preserves the communication-aware affinity established during placement. These revisions appear in \textbf{Section~\ref{sec:scheduling}} on \textbf{Page~12}, \textbf{Rows 930–960}.

\bh{Comment 2.2:}
Have the authors compared against a random placement strategy?

\bh{Response 2.2:} We agree that random placement is a useful baseline to better show the benefit of communication-aware placement. To address this comment, we added a random placement baseline to the placement evaluation in Section~\ref{sec:placmt}. Similar to the other placement policies, memory chiplets remain on the interposer outer ring. However, accelerator chiplets are placed randomly and model weights are randomly assigned to HBM chiplets. This baseline removes any intentional communication locality and serves as a lower-complexity alternative to communication-aware placement.

The new analysis shows that random placement performs similarly to round-robin placement, indicating that simply distributing resources across the interposer is insufficient to consistently reduce communication overhead. HYDRA's communication-aware placement remains consistently better. The new discussion and results added in \textbf{Section~\ref{sec:placmt}} on \textbf{Page~11}, \textbf{Rows 861–880}.




\bh{Comment 2.3:} The authors give a formula for estimating the cumulative latency of each chiplet. But how are tcomp and tcomm calculated precisely? The authors could improve the intuition on how the mixing of the workloads is done in a careful way. Are there different types of tasks that are more aligned to different types of chiplets? How are authors dealing with it? What all options did the authors consider?

\bh{Response 2.3:}
We agree that the original description should more clearly explain how $t_{\text{comp}}$ and $t_{\text{comm}}$ are obtained and how they encode task-chiplet affinity. To address this concern, we expanded Section~\ref{sec:task_sched} to clarify both latency terms and their role in scheduling.

The revised text also explains how workload mixing is handled. Each task corresponds to a phase-specific model operation, such as Attention prefill, Attention decode, SSM prefill, SSM decode, or feed-forward computation. As a result, different task types exhibit different latency profiles across chiplets. HYDRA captures these affinities through the latency tables and uses them to construct the preferred pipeline mapping. During runtime, the elastic scheduler may temporarily deviate from this preferred mapping when load imbalance outweighs the associated execution penalty. We have revised the manuscript in \textbf{Section~\ref{sec:task_sched}} on \textbf{Page~6}, \textbf{Rows 498–550}.
 

\vspace{-2mm}
\section*{\textbf{Reviewer \#3}}

\bh{Comment 3.1:}
The experiment is done on only one of the hybrid model - Nemotron-H-4B , whose layer structure is not explained in the paper. on looking up, I see that it is mostly Mamba layers with few attention layers. How would the results change when a different kind of hybrid model is used, where lets say, more attention layers are present?

\bh{Response 3.1:} 
We agree that the evaluation should better explain the hybrid model structures and include additional models with more attention layers. To address this concern, we added a new cross-model evaluation using three representative hybrid LLMs with different layer organizations: Jamba-tiny~\cite{lenz2025jamba}, Zamba2-7B~\cite{glorioso2024zamba}, and Nemotron-H-4B~\cite{blakeman2025nemotron}. The revised manuscript now explicitly reports that Jamba contains 2 Attention and 14 Mamba blocks, Zamba contains 13 Attention and 81 Mamba blocks, and Nemotron-H contains 4 Attention and 24 Mamba blocks. Although the number of Attention layers is smaller than the number of Mamba layers in all three models, \textit{Attention blocks contribute a disproportionate amount of computation, memory traffic, and KV-cache storage, making their impact on hardware design decisions significant}.

Then, we rerun \textit{HYDRA} on these three hybrid workloads and report the results obtained with the optimized configurations in a new specialization study (Table~\ref{tab:specialization_tradeoff}). 
These results show how the preferred hardware configuration changes across different hybrid model compositions and how well configurations transfer across models. At the same time, the study demonstrates that HYDRA can adapt its design recommendations to different Attention--Mamba mixtures and identify configurations that remain effective across multiple hybrid workloads.
Therefore, the revised evaluation extends the original Nemotron-H-only study and directly examines how architectural conclusions change as the composition of hybrid models varies. The detailed revisions are described in Response 1.1 (\textbf{Section~\ref{sec:tradeoff}}, on \textbf{Page 10}).
We also added a description of the hybrid models in \textbf{Section~\ref{sec:related_work}} on \textbf{Page~2}, \textbf{Rows 148–151}


\bh{Comment 3.2:}
The paper focuses mainly on throughput and TTFT. other deployment concerns such as power, thermals, and reliability are less central.

\bh{Response 3.2:}
We agree that power, thermal, and reliability are important deployment concerns. This work focuses on macro-architectural DSE for throughput and TTFT under dynamic multi-tenant workloads, leaving these additional metrics for future work. We clarify in the conclusion that HYDRA's models of chiplet placement, communication, memory utilization, and runtime scheduling provide a foundation for future power- and thermal-aware DSE.
The corresponding revisions are included in \textbf{Section~\ref{sec:conclusion}} on \textbf{Page~13}, \textbf{Rows 1034–1042}.


\bh{Comment 3.3:}
HBM is never explained in the paper.

\bh{Response 3.3:}
Thanks for pointing out this issue. The revised manuscript now defines HBM at its first use in \textbf{Section~\ref{sec:systemspec}}, on \textbf{Page 4}, \textbf{Rows 289–290}.

}{}

\ifthenelse{\boolean{CamResponse}}{
\clearpage
\section*{Response to the Post-Revision Comments}

We thank the reviewers for their careful evaluation of our revised manuscript and for recognizing that the major concerns raised in the previous review have been addressed. We also appreciate the additional minor comments, which helped us further improve the clarity and completeness of the manuscript.

In this revision, we have addressed all post-revision comments, including reporting average throughput and TTFT improvements, clarifying the sources used to obtain $t_{comp}(k,c)$, providing additional intuition for the $t_{comm}(k,c)$ formulation and its role in the framework, and noting the negligible design-time optimization overhead of random placement. All corresponding changes are highlighted in \revcam{blue} in Author's Tracked Changes submission.

\section*{\textbf{Reviewer A}}

\bh{Comment A.1:} The revision looks good to me. The new experiments were quite helpful to provide a more well-rounded evaluation. It would be useful to report average throughput and TTFT numbers in the abstract/conclusion, rather than just the peak gains.

\bh{Response A.1:} Thank you for this suggestion. We agree that reporting average improvements provides a more representative summary of \textit{HYDRA}'s performance than reporting only the peak gain. We therefore updated the related description in the Abstract, Introduction, and Conclusion Sections to report that \textit{HYDRA} achieves 1.55× the throughput and 43.7\% lower TTFT on average, while retaining the peak throughput improvement of up to 2.3$\times$. The corresponding revisions are included in 
\textbf{Abstract} on \textbf{Page~1}, \textbf{Rows 16–18}; \textbf{Section~\ref{sec:intro}}, \textbf{Page~2}, \textbf{Rows 126–127}; and \textbf{Section~\ref{sec:conclusion}}, \textbf{Page~13}, \textbf{Rows 1024–1025}.

(The revised manuscript)

\revcam{Across all workloads, HYDRA delivers 1.55$\times$ the throughput and 43.7\% lower time-to-first-token on average, with throughput gains reaching up to 2.3$\times$.}

\section*{\textbf{Reviewer B}}

\bh{Comment B.1:} The authors give the formula for t\_{comm} - some intuition there would have been better. I mean many different formulas could be used. I wonder if how much effect will it have?

\bh{Response B.1:} Thank you for pointing this out. We have expanded the discussion around $t_{comm}(k,c)$ to provide the physical intuition behind the formulation and clarify its role in \textit{HYDRA}.

The formula captures three dominant factors in NoI communication: transfer volume, communication path length, and provisioned link bandwidth. Intuitively, a larger transfer occupies NoI links for longer, while a longer path consumes link resources across more hops. Increasing the available bandwidth reduces the transfer time. We therefore use this first-order formulation as a lightweight proxy for constructing the preferred initial task mapping.  The corresponding revisions are included in \textbf{Section~\ref{sec:task_sched}} on \textbf{Page~6}, \textbf{Rows 518–521}

(The revised manuscript)

\revcam{A larger transfer occupies NoI links for longer, while a longer path consumes link resources across more hops, and higher bandwidth reduces the transfer time. This first-order model therefore provides a lightweight proxy for initial mapping.}

\vspace{2mm}
\bh{Comment B.2:} For t\_comp: The authors say “ We obtain tcomp(k,c) from our runtime measurements or estimate from published specifications for commercial chiplets [27], and published microarchitectural models for prior accelerators [3], [21]. ” — what do we mean by X or Y? Which one do they use?

\bh{Response B.2:} Thank you for pointing out this ambiguity. We have revised the manuscript to explicitly state which source is used for each chiplet in our evaluation, rather than describing several possible characterization approaches without distinguishing them. The corresponding revisions are included in \textbf{Section~\ref{sec:task_sched}} on \textbf{Page~6}, \textbf{Rows 500–505}

(The revised manuscript)

\revcam{
We select the source of $t_{\mathrm{comp}}(k,c)$ according to the chiplet being modeled. In the evaluation (Section~\ref{sec:eval}), Mamba prefill and decode latencies are derived from the microarchitectural parameters and kernel-level characterizations reported for MARCA~\cite{li2024marca}, while Attention prefill and decode latencies are derived from those reported for TSTC~\cite{liu2023tstc}.}
For commercial chiplets such as B200, \textit{HYDRA} derives latency from vendor-published specifications~\cite{nvidia_blackwell_datasheet_2024}, whereas custom designs are characterized through cycle-level RTL simulation~\cite{kanani2026duet}.

\vspace{2mm}
\bh{Comment B.3:} The study of the random placement was useful. Thank you. This scheme will have pretty much no overhead. Please do mention the low overhead side of random placement.

\bh{Response B.3:} Thank you for this suggestion. We agree that the low implementation cost of random placement is an important characteristic of this baseline. We have therefore explicitly stated that the random-placement scheme requires negligible design-time optimization overhead. The corresponding revisions are included in \textbf{Section~\ref{sec:placmt}} on \textbf{Page~11}, \textbf{Rows 872–873}

(The revised manuscript)

while the random baseline randomly places accelerator chiplets and randomly assigns model weights to HBM chiplets, \revcam{requiring negligible design-time optimization overhead.}

\section*{\textbf{Reviewer C}}

\bh{Comment C.1:} I am happy with the expanded search with new experiments on hybrid models with different attention-to-Mamba ratios (Table-iv).

\bh{Response C.1:} Thank you for the positive feedback. We are glad that the additional specialization–generality experiments and the evaluation of hybrid models with different Attention-to-Mamba ratios address the concern raised in the previous review round.

}

\ifthenelse{\boolean{setAppendix}}{

\appendix
\section{Artifact Appendix}
\label{app:artifact}

The HYDRA artifact provides the event-driven simulator, analytical accelerator
models, hybrid-LLM workload traces, hardware configurations, fast-DSE models,
curated measurements, and scripts used in the evaluation. It supports the
standard experiments associated with Sections VI-B--VI-F, including
macro-architecture DSE, the specialization--generality study, placement and
scheduling ablations, dynamic batching, and fast DSE. The standard evaluation
runs on a CPU-only Linux machine, regenerates result tables and PNG figures,
and reports the status and output location of each experiment. Expensive
exhaustive DSE sweeps are provided as optional, resumable workflows.

\subsection{Artifact Checklist}

{\small
\begin{description}
  \item[\textbf{Program:}] HYDRA event-driven simulator, analytical models,
        experiment runners, and visualization scripts.
  \item[\textbf{Compilation:}] None; the standard workflow is Python-based.
  \item[\textbf{Models:}] Llama3-7B, Mamba2-2.8B, Nemotron-H-4B,
        Jamba-tiny and Zamba2-7B hybrid-LLM configurations.
  \item[\textbf{Data sets:}] Preprocessed request-length traces from ARXIV,
        BWB, Chat, and LongWriter; no raw prompt or generated text is included.
  \item[\textbf{Run-time environment:}] Linux x86-64, Python 3.11, Bash, and
        the pinned packages in \texttt{requirements.txt}.
  \item[\textbf{Hardware:}] CPU-only; 8 CPU cores, 16\,GB memory, and
        10\,GB free disk space are recommended for the standard workflow.
  \item[\textbf{Execution:}] Automated Bash wrappers invoking Python
        simulation and post-processing scripts.
  \item[\textbf{Metrics:}] Throughput, time to first token (TTFT),
        throughput/TTFT, chiplet utilization, and fast-DSE recovery.
  \item[\textbf{Output:}] CSV result tables, PNG figures, simulator outputs,
        and per-experiment logs.
  \item[\textbf{Experiments:}] Figures 8--10 and 12, Table IV, and the
        fast-DSE evaluation in Sections VI-B--VI-F.
  \item[\textbf{Time:}] Approximately 10--20 minutes for environment setup
        and 7 minutes for the functional workflow using completed outputs.
  \item[\textbf{Publicly available:}] Yes; the repository link is given below.
  \item[\textbf{License:}] The software license is provided in the
        repository's \texttt{LICENSE} file.
  \item[\textbf{Workflow automation:}] Repository-provided Bash and Python
        scripts; no external workflow framework is required.
\end{description}
}

\subsection{Description}

\subsubsection{How to Access}

The artifact is publicly available at
\url{https://github.com/ONQLin/hydra-cases-2026-artifact}.
Clone this repository and enter its root directory before following the
instructions in \texttt{INSTALL}.

The top-level \texttt{README.md} introduces the repository, while
\texttt{REQUIREMENTS}, \texttt{INSTALL}, and \texttt{STATUS} define the
supported environment, installation test, and evaluation scope. A copy of the
accepted manuscript is included as \texttt{HYDRA\_paper.pdf}.

\subsection{Installation and Workflow}

Detailed installation and per-experiment instructions are provided in
\texttt{INSTALL} and \texttt{artifact/README.md}. The minimal setup is:

\begin{verbatim}
python3.11 -m venv .venv
source .venv/bin/activate
python -m pip install -r requirements.txt
\end{verbatim}

Run the standard functional-evaluation workflow with:

\begin{verbatim}
HYDRA_REUSE_COMPLETED=1 \
bash artifact/reproduce_all.sh
\end{verbatim}

This command runs all non-optional experiments in paper order. Individual
experiment entry points and the optional exhaustive workflows are listed in
\texttt{artifact/README.md}. The \texttt{HYDRA\_NUM\_WORKERS} environment
variable controls simulation concurrency.

\subsection{Evaluation and Expected Results}

During execution, \texttt{artifact/reproduce\_all.sh} prints the active
experiment, output directories, elapsed time, and PASS/FAIL status. A
successful evaluation ends with:

\begin{verbatim}
HYDRA reproduction completed: PASS
Figures: artifact/figures/
Data:    artifact/run_outputs/
\end{verbatim}

Generated CSV files and PNG figures are written under
\texttt{artifact/run\_outputs/} and \texttt{artifact/figures/}, respectively.
Timestamped logs and a tabular summary are written under
\texttt{output\_temp\_runs/artifact\_reproduction/}. Small machine-dependent
numerical differences do not change the reported comparisons or trends.

}

\end{document}